\documentclass[journal=jacsat,manuscript=article]{achemso}

\usepackage[version=3]{mhchem} 
\usepackage{comment}
\usepackage{xcolor}
\usepackage{siunitx}
\usepackage{bm}
\usepackage{amsmath,amssymb}
\usepackage{xr}
\usepackage[hidelinks]{hyperref}
\usepackage{nicefrac}

\makeatletter
\newcommand*{\addFileDependency}[1]{
  \typeout{(#1)}
  \@addtofilelist{#1}
  \IfFileExists{#1}{}{\typeout{No file #1.}}
}
\makeatother
\newcommand*{\myexternaldocument}[1]{%
    \externaldocument{#1}%
    \addFileDependency{#1.tex}%
    \addFileDependency{#1.aux}%
}
\myexternaldocument{supporting}

\DeclareUnicodeCharacter{03C0}{\ensuremath{\pi}}

\makeatletter

\author{Elia Turco}
\email{elia.turco@empa.ch}
\affiliation{nanotech@surfaces Laboratory, Empa -- Swiss Federal Laboratories for Materials Science and Technology, Überlandstrasse 129, 8600 D\"{u}bendorf, Switzerland}
\author{Lara Tejerina}
\altaffiliation{These authors contributed equally to this work.}
\affiliation{Department of Chemistry, University of Zurich, Winterthurerstrasse 190, 8057 Zurich, Switzerland}
\author{Gon\c{c}alo Catarina}
\altaffiliation{These authors contributed equally to this work.}
\affiliation{nanotech@surfaces Laboratory, Empa -- Swiss Federal Laboratories for Materials Science and Technology, Überlandstrasse 129, 8600 D\"{u}bendorf, Switzerland}
\author{Andres Ortega-Guerrero}
\affiliation{nanotech@surfaces Laboratory, Empa -- Swiss Federal Laboratories for Materials Science and Technology, Überlandstrasse 129, 8600 D\"{u}bendorf, Switzerland} 
\author{\\Nils Krane}
\affiliation{nanotech@surfaces Laboratory, Empa -- Swiss Federal Laboratories for Materials Science and Technology, Überlandstrasse 129, 8600 D\"{u}bendorf, Switzerland}
\author{Leo Gross}
\affiliation{IBM Research Europe -- Zurich, Säumerstrasse 4, 8803 Rüschlikon, Switzerland}
\author{Michal Juríček}
\affiliation{Department of Chemistry, University of Zurich, Winterthurerstrasse 190, 8057 Zurich, Switzerland}
\author{Shantanu Mishra}
\email{shantanu.mishra@chalmers.se}
\affiliation{IBM Research Europe -- Zurich, Säumerstrasse 4, 8803 Rüschlikon, Switzerland}
\alsoaffiliation{Department of Physics, Chalmers University of Technology, 41296 Gothenburg, Sweden}

\title{The Multiconfigurational Ground State of a Diradicaloid Characterized at the Atomic Scale
}

\begin{document}




\begin{abstract} 
{\noindent We report the tip-induced generation and scanning probe characterization of a singlet diradicaloid, consisting of two phenalenyl units connected by an sp-hybridized C\textsubscript{4} chain, on an ultrathin insulating NaCl surface. The bond-order contrast along the C\textsubscript{4} chain measured by atomic force microscopy and mapping of charge-state transitions by scanning tunneling microscopy, in conjunction with multiconfigurational calculations, reveal that the molecule exhibits a many-body ground state. Our study experimentally demonstrates the manifestation of strong electronic correlations in the geometric and electronic structures of a single molecule.}
\end{abstract}



\section{Introduction} 

Diradicaloids represent an intriguing class of compounds whose reactivity and enigmatic electronic structure have fascinated chemists and physicists for over a century\cite{stuyver_diradicals_2019,abe_diradicals_2013,wu_diradicaloids_2022}. As intermediates between diradicals and closed-shell molecules, their nature is central to understanding the chemical bond itself. Diradicaloids are represented as resonance hybrids between open- and closed-shell structures, which underscores the ambiguity in defining their electronic structure with theories that do not account for the multiconfigurational nature of electronic wavefunctions. The small gap between the highest occupied and lowest unoccupied molecular orbitals (HOMO and LUMO) of diradicaloids facilitates mixing between multiple electronic configurations in the singlet ground state. This multiconfigurational nature is commonly quantified by the diradical index, which can take values between 0 (closed-shell) and 1 (diradical).\cite{nakano_hyperpolarizability_2011,Grynova_WIRES_2025} A well-established correlation between aromatic stabilization and diradical character enables fine tuning of the electronic and magnetic properties through structural design, making diradicaloids promising candidates for applications in optoelectronics and spintronics.\cite{gopalakrishna_open-shell_2018}

Driven by progress in synthetic methods and improved characterization techniques, a number of singlet diradicaloids have been studied in recent years, for example, long acenes\cite{zuzak_-surface_2024,RuanJACS2025,HayashiChemSci2025}, periacenes\cite{sanchez-grande_unravelling_2021}, anthenes\cite{KonishiJACS2010,KonishiJACS2013,WangNatCom2016}, zethrenes\cite{HuangJACS2016,ZengAngew2016,hu_modern_2017,ZengJACS2018,turco_-surface_2021, mishra_-surface_2020}, rhombenes\cite{mishra_large_2021, biswas_steering_2023} and indenofluorenes\cite{ShimizuAngew2013,DresslerAngew2017,DiGiovannantonioJACS2019,MishraNatChem2024}. The advent of on-surface chemistry\cite{ClairChemRev2019} has extended the scope of the study of singlet diradicaloids to the atomic scale by means of scanning probe techniques. In this context, it is important to understand how the multiconfigurational ground state of singlet diradicaloids manifests in terms of the fundamental observables of a molecular system such as bond order and molecular orbital densities.

\begin{figure}[t]
    \centering
    \includegraphics[width=\linewidth]{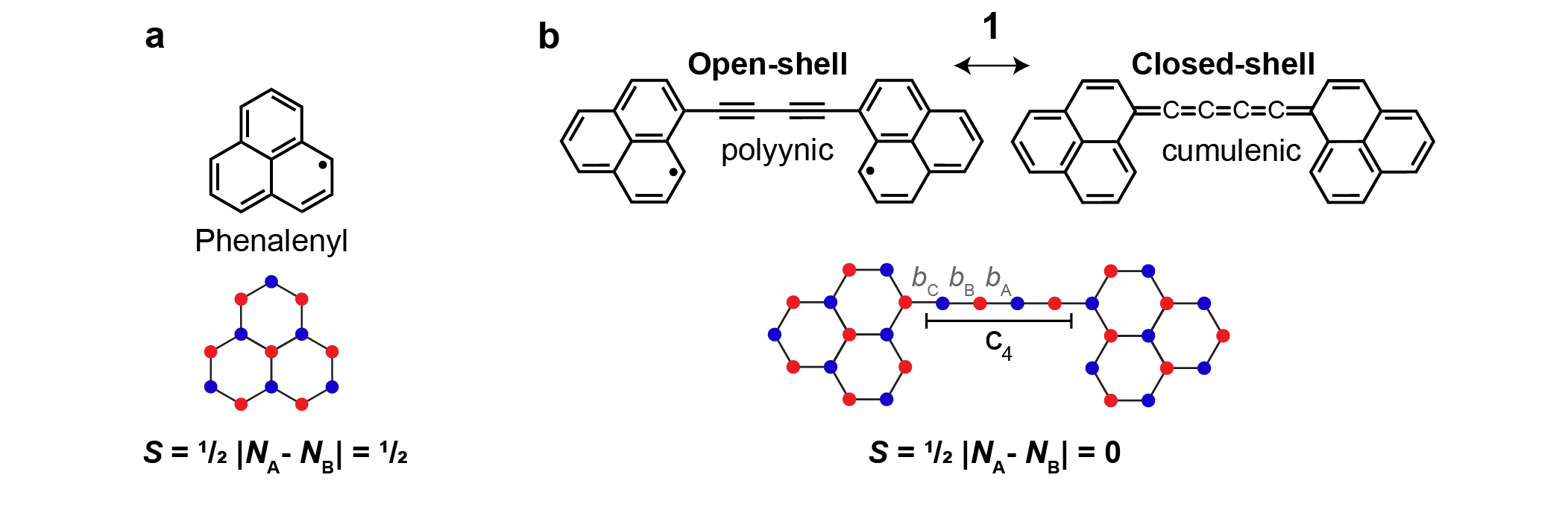}
    \caption{(a, b) Chemical structures and sublattice representations of phenalenyl radical (a) and compound \textbf{1} (b). The two sublattices are represented with different colors. For \textbf{1}, two possible resonance structures are shown, namely, an open-shell with a polyynic C\textsubscript{4} chain, and a closed-shell with a cumulenic C\textsubscript{4} chain. For the closed-shell resonance structure, the bond orders of \textit{b}\textsubscript{A}, \textit{b}\textsubscript{B} and \textit{b}\textsubscript{C} should be similar, while for the open-shell resonance structure, \textit{b}\textsubscript{B} should have a higher bond order than \textit{b}\textsubscript{A} and \textit{b}\textsubscript{C}.} 
    \label{fig:scheme}
\end{figure}

In this study, we use atomic force microscopy (AFM) and scanning tunneling microscopy (STM) to generate and study a diradicaloid \textbf{1} (C\textsubscript{30}H\textsubscript{16}, Fig. \ref{fig:scheme}), and elucidate the influence of electronic correlations on the geometric and electronic structures of the molecule. Compound \textbf{1} consists of two phenalenyl units (Fig. \ref{fig:scheme}a), sp\textsuperscript{2}-conjugated polycyclic conjugated hydrocarbons with an \textit{S} = 1/2 (doublet) ground state (\textit{S} denotes the total quantum spin number), which are connected through their majority sublattice carbon atoms via an sp-hybridized C\textsubscript{4} chain, resulting in an \textit{S} = 0 (singlet) ground state as per Ovchinnikov’s rule\cite{ovchinnikov_multiplicity_1978_capitalized, Elliot_Lieb_1989}.

As shown in Fig. \ref{fig:scheme}b, compound \textbf{1} can be represented as a resonance hybrid of two structures, namely, open- and closed-shell singlets. Importantly, the two resonance structures present different bonding motifs in the C\textsubscript{4} chain, namely, polyynic (alternating single and triple bonds) and cumulenic (all double bonds) for the open- and closed-shell structures, respectively. Previously, Hirao et al. studied a derivative of \textbf{1} in single-crystalline form.\cite{hirao_spinspin_2021_int} Upon cooling the sample from 250 to 100 K, the authors observed a slight increase (decrease) in the bond lengths of the bonds labeled \textit{b}\textsubscript{A} and \textit{b}\textsubscript{C} (\textit{b}\textsubscript{B}) in Fig. \ref{fig:scheme}b. This can be interpreted as an enhancement of the diradical character, that is, increasing contribution from the open-shell resonance form, as the temperature is lowered. The different bonding motifs for the two resonance structures make \textbf{1} a suitable system for characterization by AFM, which can distinguish C--C bonds of different bond orders~\cite{gross_bond-order_2012}. We show by AFM imaging that compared to a polyynic bonding motif with formal C--C single and triple bonds, the C\textsubscript{4} chain in \textbf{1} exhibits a markedly reduced bond-order contrast. Furthermore, STM imaging of \textbf{1} at the ion resonances reveals orbital densities that cannot be explained on the basis of charge-state transitions involving a single-determinant ground state of \textbf{1} (such as a closed-shell configuration with doubly occupied HOMO and empty LUMO). However, the experimental results can be explained well if one considers a ground state of \textbf{1} that is composed of multiple Slater determinants, that is, a multiconfigurational ground state.

\section{Results and Discussion} 
\subsection{Generation and Structural Characterization}
\begin{figure}[h!]
    \centering
    \includegraphics[width=\linewidth]{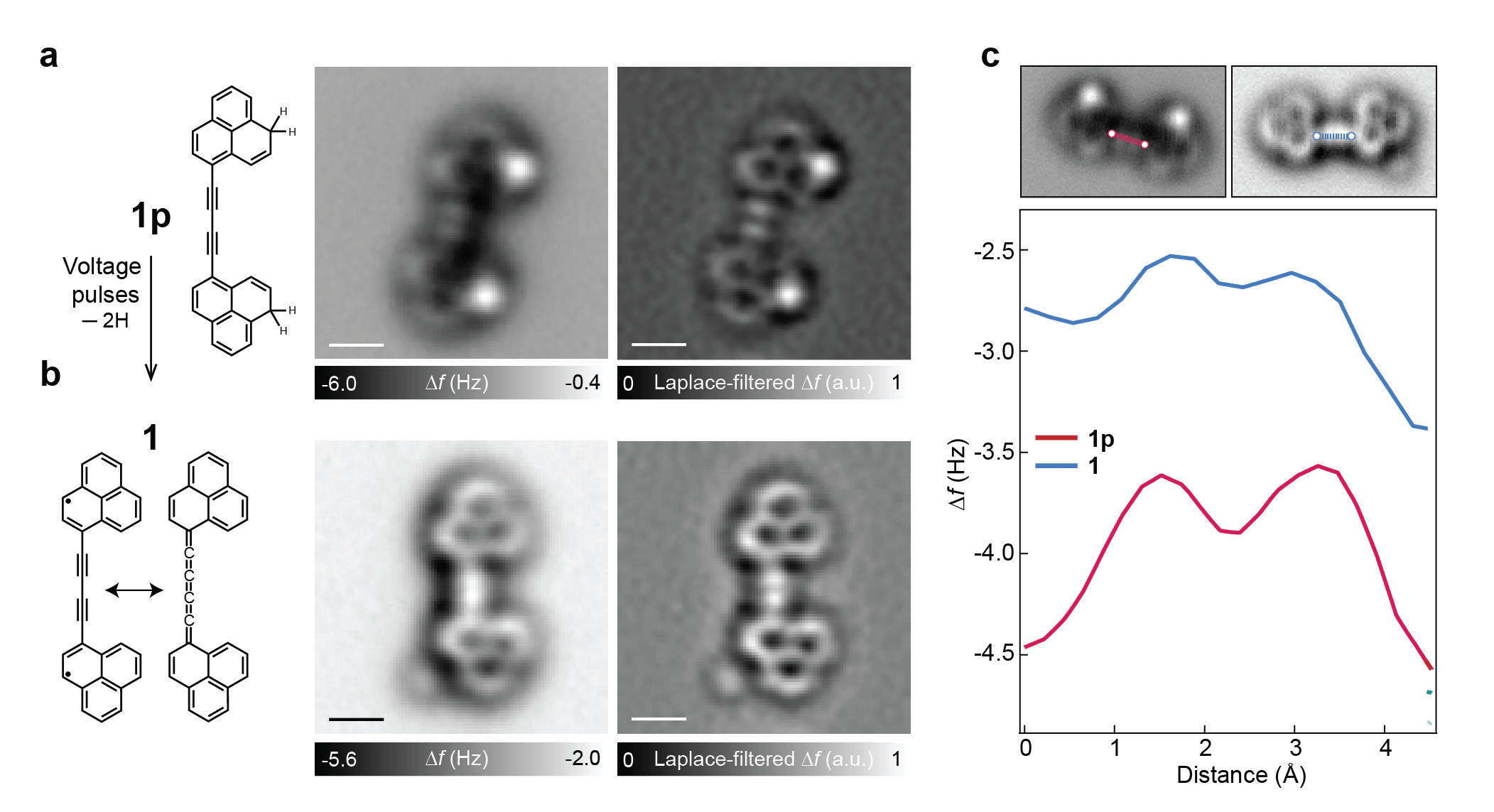}
    \caption{Structural characterization of \textbf{1p} and \textbf{1}. (a, b) From left to right: chemical structures, AFM images and corresponding Laplace-filtered AFM images of \textbf{1p} (a) and \textbf{1} (b); STM set-point: $V = 0.2$~V and $I = 1.0$~pA on NaCl, tip height $\Delta z=0.5$ \AA{}. a.u. denotes arbitrary units. (c) $\Delta f$ line profiles along the C\textsubscript{4} chains of \textbf{1p} (red) and \textbf{1} (blue). Scale bars: 0.5 nm.}
    \label{fig:AFM}
\end{figure}

Compound \textbf{1} was generated from the corresponding dihydro precursor \textbf{1p} (Fig. \ref{fig:AFM}a), which was synthesized in solution (Methods and Figs. S1--S8). A submonolayer coverage of \textbf{1p} was sublimed on a Cu(111) surface partially covered by bilayer NaCl films (Fig. S9). STM and AFM imaging showed the coexistence of both \textit{cis} and \textit{trans} isomers of \textbf{1p}, which differ in the relative orientation of the two phenalene units through rotation via the C--C single bond connecting the phenalene units and the C\textsubscript{4} chain. AFM imaging (Figs. \ref{fig:AFM} and S10) revealed that both isomers adopt a mostly planar geometry on NaCl, and the sp-hybridized C\textsubscript{4} chain exhibits a polyynic bonding motif evidenced by a modulation of the frequency shift ($\Delta f$) signal along the chain, as described below. Compound \textbf{1} was generated by applying voltage pulses to individual \textbf{1p} molecules by the tip of the STM/AFM system, which led to homolytic cleavage of the two C($sp^3$)-H bonds\cite{turco_observation_2023,PavlicekNatNano2017}. The sequential manipulation of \textbf{1p} to \textbf{1} was monitored by AFM imaging, with an example shown in Fig. S11.

We now focus on elucidating the difference in the bond-order contrasts of the C\textsubscript{4} chain in \textbf{1p} and \textbf{1}. In AFM imaging, chemical bonds with higher bond orders show a larger $\Delta f$ signal due to stronger repulsive forces. Figure~\ref{fig:AFM}a shows an AFM image of \textbf{1p}, where two bright features that correspond to the formal C--C triple bonds\cite{de_oteyza_direct_2013,pavlicek_polyyne_2018} labeled \textit{b}\textsubscript{B} in Fig. \ref{fig:scheme}b (that exhibit a higher bond order than the neighboring formal C--C single bonds labeled \textit{b}\textsubscript{A} and \textit{b}\textsubscript{C} in Fig. \ref{fig:scheme}b) are evident at the center of the C\textsubscript{4} chain. Note that the two bright features at the phenalene units correspond to the dihydro groups. Compared to \textbf{1p}, there is a noticeable decrease in the bond-order contrast of the C\textsubscript{4} chain in AFM imaging of \textbf{1} (Fig.~\ref{fig:AFM}b, see also Figs. S11 and S12). This difference in the bond-order contrast is also visualized in the $\Delta f$ line profiles along the C\textsubscript{4} chains of \textbf{1p} and \textbf{1} (Fig. \ref{fig:AFM}c), where \textbf{1p} exhibits a larger $\Delta f$ modulation compared to \textbf{1}. However, the fact that a $\Delta f$ modulation remains in \textbf{1} indicates that the C\textsubscript{4} chain in \textbf{1} is neither polyynic with formal C--C single and triple bonds (as in \textbf{1p}) nor cumulenic (where no bond order contrast should be visible\cite{sun_-surface_2023}).

\subsection{Electronic characterization}
Based on AFM imaging of \textbf{1} that reveals bond-order contrast in the C\textsubscript{4} chain that is intermediate between polyynic (corresponding to a diradical state with two singly occupied molecular orbitals) and cumulenic (corresponding to a closed-shell state with doubly occupied HOMO and empty LUMO) geometries, and the small DFT-calculated HOMO-LUMO gap of \textbf{1} (Fig. S13), it is likely that the system exhibits a multiconfigurational ground state. In line with this expectation, STM imaging of \textbf{1} at voltages corresponding to the ion resonances reveals orbital densities that cannot be explained with a single-reference picture but requires invoking a multiconfigurational framework, as discussed below. Here, tunneling events at the ion resonances are considered as many-body transitions between different charge states of \textbf{1}, and the electronic ground state of \textbf{1} is described by weighted combinations of multiple Slater determinants.

\begin{figure}[h!]
    \centering
    \includegraphics[width=\linewidth]{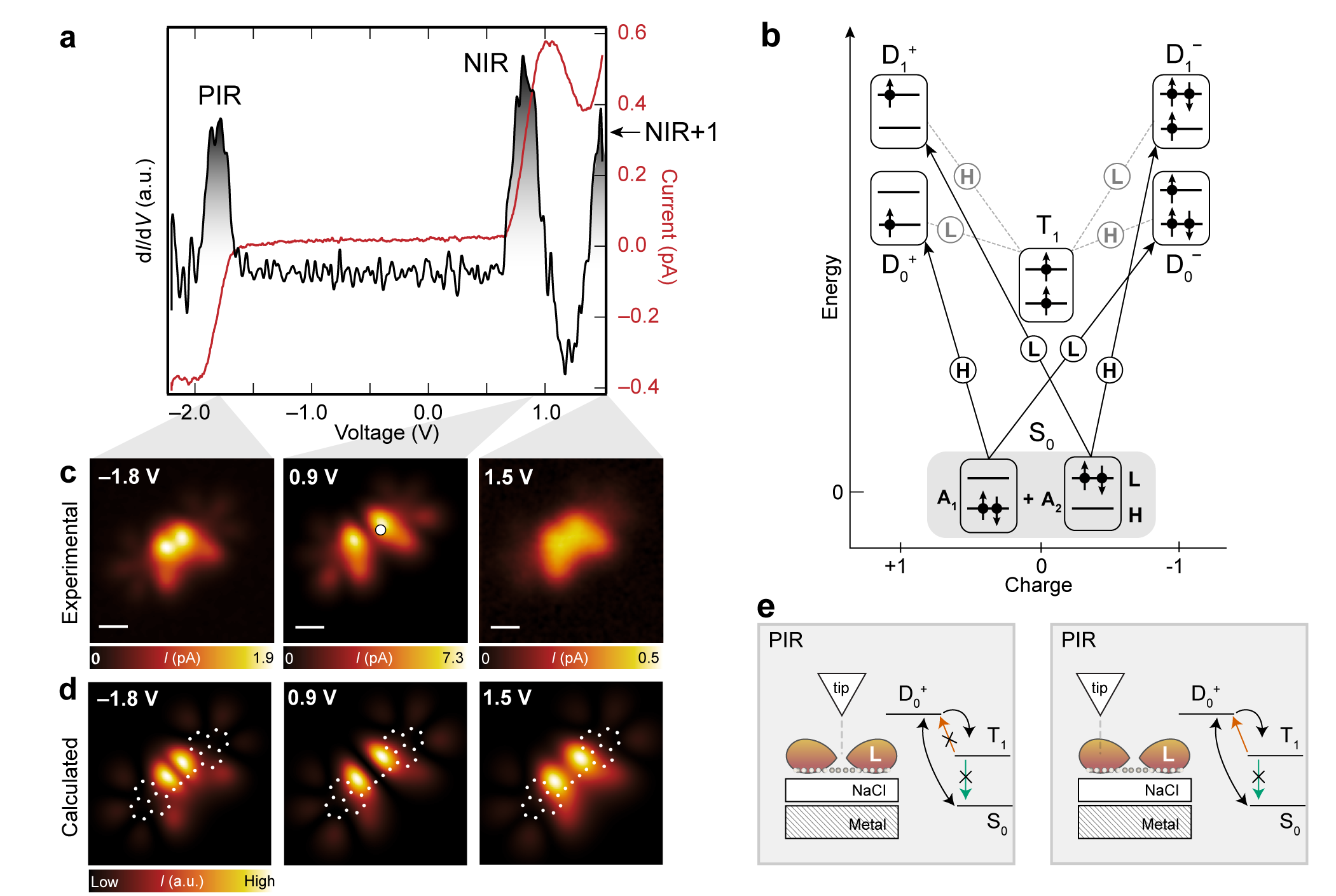}
    \caption{Electronic characterization of \textbf{1}. (a) Constant-height $I(V)$ spectrum and the corresponding d\textit{I}/d\textit{V}(\textit{V}) spectrum acquired on \textbf{1} (open feedback parameters on the molecule: $V = -2.2$ V, $I = 0.4$ pA). The acquisition position is indicated by the filled white circle in (c). PIR and NIR denote the positive and negative ion resonances, respectively. (b) Scheme of the many-body transitions relevant for the measured ion resonances. The transitions are labeled according to the orbital involved: \textbf{H} (HOMO) or \textbf{L} (LUMO). The $S_0 \rightarrow D_1^+$ transition was not accessible in the experimental voltage range. (c) Constant-height STM images at the ion resonances (from left to right, open feedback parameters on NaCl: $V =-1.8$ V, $I = 1.0$~pA, $\Delta z = 1.9$ \AA{}; $V = 0.9$~V,  $I = 1.0$~pA, $\Delta z = 2.3$ \AA{};  $V = 1.5$ V, $I = 1.0$~pA, $\Delta z = 3.5$ \AA{}). (d) Transition probability maps at the voltages corresponding to the experimental images in (c) (see Figs. S18, S22 and S23). The molecular structure of \textbf{1} is overlaid on the maps. (e) Sketch illustrating triplet trapping for tip positions above the LUMO nodal plane (left), and no trapping for tip positions above finite LUMO density (right). The data in (a, c) were acquired with a metallic tip. Scale bars: 0.5 nm.}
    \label{fig:Electronic structure}
\end{figure}

Figure~\ref{fig:Electronic structure}a (see also Figs. S14--S17) shows a differential conductance spectrum (d\textit{I}/d\textit{V}(\textit{V}), where \textit{I} and \textit{V} denote the tunneling current and bias voltage, respectively) acquired on \textbf{1} exhibiting three peaks centered at --1.8, 0.9 and 1.5 V, labeled PIR (positive ion resonance), NIR (negative ion resonance) and NIR+1, respectively. In a single-reference picture and assuming a closed-shell electronic configuration, the peak at --1.8 V should correspond to electron detachment from the HOMO of \textbf{1}. As shown in Fig. S18, the calculated HOMO local density of states (LDOS) map exhibits a maxima at the center of the C\textsubscript{4} chain. Although the STM image at --1.8 V (Fig. \ref{fig:Electronic structure}c) shows a high intensity at the center of the C\textsubscript{4} chain, there is a concomitant depression reminiscent of a nodal plane, which is not explained by a transition involving only the HOMO. At positive biases, the resonance at 0.9 V should correspond to electron attachment to the LUMO of \textbf{1}, and in this case, the LUMO LDOS map (Fig. S18) agrees well with the STM image at 0.9 V. However, for the resonance at 1.5 V, which should correspond to electron attachment to the LUMO+1 of the molecule, the corresponding STM image should show the superposition of LUMO and LUMO+1 densities (because electron attachment to both LUMO and LUMO+1 is possible at this bias). The STM image at 1.5 V does not agree with the LDOS map corresponding to the superposition of the LUMO and LUMO+1 (Fig. S18), but counterintuitively, agrees better with the LDOS map corresponding to the superposition of the HOMO and LUMO. Clearly, a single-reference picture fails to account for these peculiar features in the STM images.

The experimental STM images can be reconciled with measured orbital densities if one instead considers a multiconfigurational picture, as demonstrated previously for a similar case.\cite{yu_apparent_2017} Within a minimal multiconfigurational framework, the neutral singlet ground state of \textbf{1} ($S_0$) is described as a linear combination of two Slater determinants $\psi_B$ and $\psi_{AB}$,
\begin{equation}
    S_0 = A_1 \psi_B + A_2 \psi_{AB}
    \label{Eq: GS}
\end{equation}
where $\psi_B = |\uparrow \downarrow \rangle_{\mathrm{HOMO}} |0 \rangle_{\mathrm{LUMO}}$ and $\psi_{AB} = |0 \rangle_{\mathrm{HOMO}} |\uparrow \downarrow \rangle_{\mathrm{LUMO}}$ correspond to electronic configurations with bonding (doubly occupied HOMO and empty LUMO) and anti-bonding (empty HOMO and doubly occupied LUMO) symmetries, respectively. 
To validate our assumption about the multiconfigurational ground state of \textbf{1} (Eq.~\ref{Eq: GS}) we performed calculations using Hubbard model density matrix renormalization group (DMRG), as well as complete active space self-consistent field (CASSCF) (Methods, Figs. S13 and S19--S22, and Tables S1 and S2). All calculations were done for three distinct geometries of \textbf{1}:
a DFT PBE0-XC UKS optimized structure, a cumulenic structure with all double bonds (from PBE0 optimized ethylene) in the C\textsubscript{4} chain, and a polyynic geometry with alternating single and triple bonds (from PBE0 optimized propyne) in the C\textsubscript{4} chain.
For brevity, we will focus on the DMRG results, yielding the most accurate description\cite{catarina_conformational_2024} of the multiconfigurational nature of \textbf{1}.
The calculations corroborate the assumption of a singlet ground state with a predominantly bonding character ($|A_1|^2 / |A_2|^2 > 4$) across all considered geometries. The weight of the doubly excited configuration, $|A_2|^2$, ranges from 0.04 to 0.12, increasing from the cumulenic to the polyynic structure. For the DFT PBE0-XC UKS optimized structure, which is the focus of the following analysis, we obtain $|A_1|^2=0.60$ and $|A_2^2|=0.06$. Since $|A_1|^2 + |A_2|^2 < 1$, the ground state $S_0$ involves more Slater determinants than suggested by the simplified picture of Eq. \ref{Eq: GS}. However, as we show, this two-configuration picture describes well the observed electronic properties of \textbf{1}. 

%


Figure~\ref{fig:Electronic structure}b illustrates a schematic of the many-body electronic transitions corresponding to the measured ion resonances of \textbf{1}. To elucidate the excitation mechanisms responsible for the experimental STM images in Fig.~\ref{fig:Electronic structure}c, we modeled the system using a master equation (Methods, Fig. S23 and Table S3) that incorporates all relevant tunneling pathways and a finite lifetime of the neutral excited states. The spatially resolved transition probability maps (Fig.~\ref{fig:Electronic structure}d) associated with the transitions depicted in Fig.~\ref{fig:Electronic structure}b were derived from Dyson orbitals computed via DMRG (Figs. S18, S22 and S23). For this, we assumed tunneling with an s-wave tip, because of the predominant s-wave tunneling character at large tip-sample distances, even for carbon monoxide-functionalized tips\cite{paschke_distance_2025}.
Starting from $S_0$, we assign the positive ion resonance at --1.8 V to a resonant transition to the cationic doublet ground state $D_0^+$, wherein an electron is detached from the HOMO. The system subsequently decays to $S_0$ by electron transfer from the surface, resulting in a net current. However, the corresponding $S_0 \rightarrow D_0^+$ Dyson orbital (Fig. S22) does not feature a central nodal plane, as observed in the STM image at --1.8 V. To resolve this discrepancy, one must consider the role of the neutral triplet excited state $T_1$. Following an initial $S_0 \rightarrow D_0^+$ resonant tunneling event, the system can be neutralized via electron transfer from the surface in two different ways. By electron attachment to the HOMO, the system decays to $S_0$, while by electron attachment to the LUMO, the system decays to $T_1$ (located $\sim$0.36 eV above $S_0$ according to DMRG calculations). If the system is in $T_1$, it can either decay to $S_0$ (which is a slow process because it requires a change in the spin multiplicity), or the system can be excited to $D_0^+$ via electron tunneling to the tip, followed by a decay to $S_0$. As shown schematically in Fig. ~\ref{fig:Electronic structure}e, the amplitude of the $T_1 \rightarrow D_0^+$ transition, which involves electron detachment from the LUMO, is strongly dependent on the tip position. If the tip is located at the center of the C\textsubscript{4} chain (where the LUMO exhibits a nodal plane) the system is trapped in $T_1$. The tunneling channel through the molecule is effectively blocked and the total current is reduced at the nodal plane, as observed in the STM image. Away from the chain center, where the LUMO has a non-zero amplitude, this transition can take place. The corresponding transition probability map, which takes into account the $S_0 \rightarrow D_0^+$ and $T_1 \rightarrow D_0^+$ transitions (Fig. ~\ref{fig:Electronic structure}d), exhibits good agreement with the STM image at --1.8 V. We note that transitions involving ground and excited states of a molecule have been previously predicted for copper phthalocyanine,\cite{siegert_nonequilibrium_2016} and experimentally observed for several molecular systems by STM/AFM-based spectroscopy\cite{yu_apparent_2017,FatayerPRL2021,PengScience2021,Sellies2025NatNano} and STM-induced luminescence measurements\cite{JiangPRL2023,HungPRR2023}.

The first negative ion resonance at 0.9 V represents a resonant transition from $S_0$ to the anionic doublet ground state $D_0^-$, corresponding to electron attachment to the LUMO. The Dyson orbital for this transition features the characteristic central nodal plane as seen in the experiment. The system may subsequently decay to either $S_0$ or $T_1$, as described above. The decay to $T_1$ opens an additional tunneling channel via the $T_1 \rightarrow D_{0}^-$ transition, which does not feature a central nodal plane (as it involves electron attachment to the HOMO). However, this pathway does not alter the appearance of the STM image because if the tip is positioned above the nodal plane of the $S_0 \rightarrow D_{0}^-$ transition, the system is not excited to $D_0^-$ and, therefore, cannot decay to $T_1$. The transition probability map that takes into account the $S_0 \rightarrow D_{0}^-$ and $T_1 \rightarrow D_{0}^-$ transitions reproduces the experimental STM image at 0.9 V. The second negative ion resonance at 1.5 V corresponds to a resonant transition from $S_0$ to the anionic doublet excited state $D_1^-$, where an electron is attached to the HOMO. This process becomes possible because of the multiconfigurational ground state of \textbf{1}, where the 
$\psi_{AB}$ component of $S_0$ contributes. The transition probability map, which, besides the resonant $S_0 \rightarrow D_{1}^-$ transition, includes contributions from the off-resonant $S_0 \rightarrow D_0^-$ transition (accessible at 1.5 V but with reduced spectral weight), and the $T_1 \rightarrow D_{0}^-$ and $T_1 \rightarrow D_{1}^-$ transitions, exhibits good agreement with the experimental STM image at 1.5 V.

The explanations for the effects of the multiconfigurational ground state on STM orbital density images were brought forward by Yu et al.\cite{yu_apparent_2017} Here, in addition to the STM orbital density images, we also observe signatures of the multiconfigurational ground state of a molecule by AFM, revealing contributions from both cumulenic and polyynic resonance structures.



\section{Conclusions}

We presented generation and characterization of a neutral diradicaloid \textbf{1}, revealing experimental signatures and observables related to its multiconfigurational ground state in scanning probe measurements. We showed that the maps of charge-state transitions of \textbf{1} measured by STM cannot be explained by a picture wherein electron detachment or attachment takes place in the framework of single-particle states; but can only be explained if the ground state of \textbf{1} is considered to be a multiconfigurational state consisting of weighted combinations of multiple Slater determinants. Moreover, and in line with the observations of the STM measurements, AFM imaging reveals that the C\textsubscript{4} bridge of \textbf{1} exhibits a bond-order contrast that is intermediate between polyynic and cumulenic bonding motifs, lending support to the picture that \textbf{1} is neither a diradical nor a closed-shell system, but a diradicaloid best described as a resonance hybrid of open- and closed-shell states. Our study thus provides a striking example of strong electronic correlations manifesting at the atomic scale in real space.

\begin{acknowledgement}
E.T. would like to thank Carlo Antonio Pignedoli and Aliaksandr Yakutovich for fruitful discussions. 
This research was financially supported by the H2020-MSCA-ITN (ULTIMATE, No. 813036); the Swiss National Science Foundation (SNF-PiMag, grant numbers CRSII5\_205987, 212875, PP00P2\_170534, and TMCG-2\_213829/CASCADER); the Werner Siemens Foundation (CarboQuant); and the European Research Council Synergy Grant MolDAM (grant no. 951519). A.O-G and G.C. acknowledge financial support from the NCCR MARVEL, a National Centre of Competence in Research funded by the Swiss National Science Foundation (grant number 205602). L.T. and M.J. also acknowledge the Dr. Helmut Legerlotz Foundation (University of Zurich).

\end{acknowledgement}


\begin{suppinfo}

Experimental and theoretical methods, additional scanning probe and theoretical data, and solution synthesis and characterization. The raw NMR data are available free of charge on the public repository Zenodo under the link: https://zenodo.org/record/15311165 (DOI: 10.5281/zenodo.15311165).

\end{suppinfo}


\bibliography{refs_all}
\end{document}


\newpage
\newpage
\section{Methods: solution synthesis and characterization}

\textbf{General information.} Glassware utilized in the reactions, carried out under both anhydrous and non-anhydrous conditions, were cleaned and dried in an oven at 150~\textdegree{}C for at least 24 h prior to the experiment. All reagents and solvents, including non-anhydrous and anhydrous solvents such as \ce{CH2Cl2}, cyclohexane, EtOAc or THF, were supplied from commercial sources and used without additional purification unless otherwise noted. Thin-layer chromatography (TLC) was used to monitor the reactions, using aluminium sheets covered with silica gel containing fluorescent indicator UV254 (available from Alugram SIL G/UV254, Macherey-Nagel or Sigma-Aldrich) and viewed under UV light (254 or 365 nm). Silica gel 60 (230–400 mesh, Sigma-Aldrich) was used for flash column chromatography.

All $^1$H and $^{13}$C NMR spectra were recorded at 25~\textdegree{}C on a Bruker 400 MHz spectrometer. Chemical shifts ($\delta$) are reported in parts per million (ppm) relative to the solvent residual peak: CDCl$_3$ ($\delta = 7.26$ ppm for $^1$H and 77.16 ppm for $^{13}$C)\cite{fulmer_nmr_2010}.

Electrospray ionisation (ESI) high-resolution mass spectra (HRMS) were recorded on a timsTOF Pro TIMS-QTOF-MS instrument (Bruker Daltonics GmbH, Bremen, Germany). The samples were dissolved (e.g., in MeOH) at a concentration of approximately 50~$\mu$g~mL$^{-1}$ and analyzed via continuous flow injection (2~$\mu$L~min$^{-1}$). The mass spectrometer was operated in the positive (or negative) electrospray ionization mode at 4,000 V (–4,000 V) capillary voltage and –500 V (500 V) endplate offset with a nitrogen nebulizer pressure of 0.4 bar and a dry gas flow of 4 L~min$^{-1}$ at 180~\textdegree{}C. Mass spectra were acquired in a mass range from m/z 50 to 2,000 at approximately 20,000 resolution (m/z 622) and at 1.0 Hz rate. The mass analyzer was calibrated between m/z 118 and 2,721 using an Agilent ESI-L low concentration tuning mix solution (Agilent, USA) at a resolution of 20,000, giving a mass accuracy below 2 ppm. All solvents used were purchased in best LC-MS quality.

\begin{figure}
    \centering
    \includegraphics[width=0.5\linewidth]{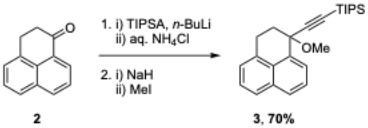}
    \caption{\\ Synthesis of triisopropyl((1-methoxy-2,3-dihydro-1H-phenalen-1-yl)ethynyl)silane (\textbf{3}).}
    \label{fig:enter-label}
\end{figure}

The reaction was carried out under inert conditions. To a cooled (–78~\textdegree{}C) solution of (triisopropylsilyl)acetylene (TIPSA; 0.99 mL, 4.4 mmol) in THF (5 mL), \emph{n}-BuLi (2.2 mL, 3.5 mmol, 1.6 M in hexanes) was added dropwise over 10 min and the reaction mixture was stirred at –78~\textdegree{}C for 2 h. Then, a solution of 2,3-dihydro-1H-phenalen-1-one (2; 0.20 g, 1.1 mmol) in THF (6 mL) was added dropwise at –78~\textdegree{}C, and the mixture was stirred for 15 min before it was allowed to warm up to room temperature and stirred for additional 2 h. The reaction mixture was poured in ice-cold aq. \ce{NH4Cl} (sat.) and extracted with diethyl ether (3 × 40 mL). The combined organic layers were dried over anhydrous \ce{MgSO4} and filtered, and the solvent was evaporated under vacuum. The residue was passed through a silica plug (\ce{CH2Cl2}) to yield a hydroxy intermediate as a colorless oil that was used in step 2 without further purification.

To a solution of the hydroxy intermediate dissolved in THF (9 mL), \ce{NaH} (0.28 g, 6.9 mmol, 60\% dispersion in mineral oil) was added. The reaction mixture was stirred for 30 min at room temperature before methyl iodide (0.61 g, 0.27 mL, 4.3 mmol) was added and the reaction mixture was stirred at 40~\textdegree{}C for 16 h. Then, the mixture was filtered through a celite plug and the solvent was evaporated under vacuum. The residue was purified by column chromatography (SiO$_2$, cyclohexane to cyclohexane/ethyl acetate 200:1) to afford the desired product as a yellow-brown oil (293 mg, 0.774 mmol) in 70\% yield over the two steps.

\textbf{$^1$H NMR} (400 MHz, CDCl$_3$, ppm): $\delta$ 7.89 (dd, J = 7.1, 1.2 Hz, 1H), 7.76 (dd, J = 8.3, 1.2 Hz, 1H), 7.63 (d, J = 8.2 Hz, 1H), 7.41 (dd, J = 8.3, 7.1 Hz, 1H), 7.33 (dd, J = 8.2, 7.0 Hz, 1H), 7.23 (d, J = 7.0 Hz, 1H), 3.40 (ddd, J = 16.9, 12.8, 4.7 Hz, 1H), 3.27 (s, 3H), 2.95 (ddd, J = 16.3, 4.0, 3.9 Hz, 1H), 2.57 (ddd, J = 13.4, 4.7, 3.3 Hz, 1H), 2.30 (ddd, J = 13.1, 13.1, 4.6 Hz, 1H), 1.08–1.03 (m, 21H).

\textbf{$^{13}$C NMR} (101 MHz, CDCl$_3$): $\delta$ 134.8, 134.0, 133.3, 129.1, 128.2, 126.2, 125.7, 125.6, 124.93, 124.91, 107.7, 88.1, 74.3, 51.4, 35.5, 25.5, 18.8, 11.4.

\textbf{HRMS} (ESI): m/z: [M + Na]$^+$ Calcd for C$_{25}$H$_{34}$OSi 401.2271; Found 401.2265, [M + K]$^+$ Calcd for C$_{25}$H$_{34}$OSi 417.2011; Found 417.2005.

\begin{figure}
    \centering
    \includegraphics[width=0.7\linewidth]{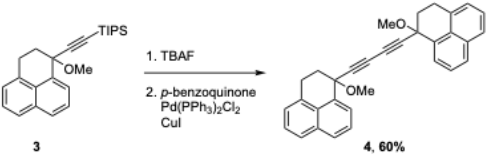}
    \caption{1,4-Bis(1-methoxy-2,3-dihydro-1H-phenalen-1-yl)buta-1,3-diyne (\textbf{4}).}
    \label{fig:enter-label}
\end{figure}

The reaction was carried out under inert conditions. To a solution of compound \textbf{3} (293 mg, 0.774 mmol) in dichloromethane (30 mL), TBAF (4.64 mL, 4.64 mmol, 1 M solution in THF) was added at room temperature and the reaction mixture was stirred for 30 min before it was filtered through a silica plug (\ce{CH2Cl2}) and the solvent was evaporated under vacuum. The alkyne intermediate was used immediately in step 2 without further purification.

To a solution of the alkyne intermediate in toluene (50 mL), diisopropylamine (1 mL) was added and the mixture was stirred for 15 min. Then, \ce{Pd(PPh3)2Cl2} (5.44 mg, 7.75 $\mu$mol), \ce{CuI} (14.8 mg, 77.5 $\mu$mol), and \emph{p}-benzoquinone (33 mg, 0.31 mmol) were added, and the reaction mixture was stirred at 50~\textdegree{}C for 2 h. The solvent was evaporated under vacuum and the residue was purified by column chromatography (SiO$_2$, cyclohexane to cyclohexane/ethyl acetate 100:1) to afford the desired product as a red solid (101 mg, 0.228 mmol) in 60\% yield over the two steps.

\textbf{$^1$H NMR} (400 MHz, CDCl$_3$, ppm): $\delta$ 7.90 (dd, J = 7.2, 1.2 Hz, 2H), 7.86 (dd, J = 8.3, 1.3 Hz, 2H), 7.73 (d, J = 8.2 Hz, 2H), 7.51 (dd, J = 8.3, 7.1 Hz, 2H), 7.43 (dd, J = 8.2, 7.0 Hz, 2H), 7.33 (dd, J = 6.9, 1.3 Hz, 2H), 3.44 (ddd, J = 16.3, 11.5, 4.6 Hz, 2H), 3.37 (s, 6H), 3.10 (ddd, J = 16.4, 4.6, 4.6 Hz, 2H), 2.59 (ddd, J = 13.1, 4.6, 4.6 Hz, 2H), 2.41 (ddd, J = 13.1, 11.6, 4.6 Hz, 2H).

\textbf{$^{13}$C NMR} (101 MHz, CDCl$_3$): $\delta$ 134.4, 134.0, 132.8, 129.4, 128.0, 126.3, 125.8, 125.6, 125.1, 125.0, 80.2, 74.6, 71.2, 51.9, 34.8, 25.7.

\textbf{HRMS} (ESI): m/z: [M + Na]$^+$ Calcd for C$_{32}$H$_{26}$O$_2$ 465.1825; Found 465.18304.

\begin{figure}[h!]
    \centering
    \includegraphics[width=0.7\linewidth]{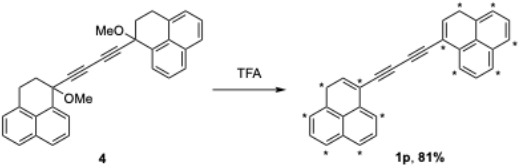}
    \caption{1,4-Di(1H-phenalen-3-yl)buta-1,3-diyne (\textbf{1p}).}
    \label{fig:enter-label}
\end{figure}

The reaction was performed under inert conditions. Compound \textbf{4} (16 mg, 36 $\mu$mol) was dissolved in argon-saturated CDCl$_3$ (0.6 mL) in an NMR tube. Trifluoroacetic acid (TFA; 4.0 $\mu$L, 54 $\mu$mol) was then added, and the tube was gently shaken at room temperature, leading to an immediate darkening of the solution. The reaction progress was monitored by $^1$H NMR spectroscopy. The signals corresponding to compound \textbf{4} disappeared within 25 min; however, the reaction was monitored for a total of 2 h. The solvents, including methanol (formed as a side product), were evaporated directly from the NMR tube to afford the desired product (11 mg, 29 $\mu$mol, 81\%) as a solid. The product was stored under argon in the original NMR tube and kept in the freezer. The sample was only freshly opened prior to surface studies. Several attempts to purify the compound via short-column chromatography (SiO$_2$ or Al$_2$O$_3$) under inert conditions were unsuccessful, as the product decomposed during the process. Due to its high sensitivity, high-resolution mass spectrometry (HRMS) could not be successfully performed.

\textbf{$^1$H NMR} (400 MHz, CDCl$_3$): Compound \textbf{1p} is obtained as a mixture of regioisomers, which differ by positions of the methylene groups that can occupy any $\alpha$-position (marked with asterisks) of the phenalenyl subunit. This mixture of regioisomers gives a complex $^1$H NMR spectrum with characteristic signals for the methylene groups ($\sim$4 ppm). The ratio between the integrated signal intensity of all methylene groups and all aromatic protons is $\sim$2:7, matching the expected value.

\vspace{0.5em}
\noindent

\noindent
\textbf{Copies of NMR spectra:}
\vspace{1em}

\begin{figure}[h!]
    \centering
    \includegraphics[width=.6\linewidth]{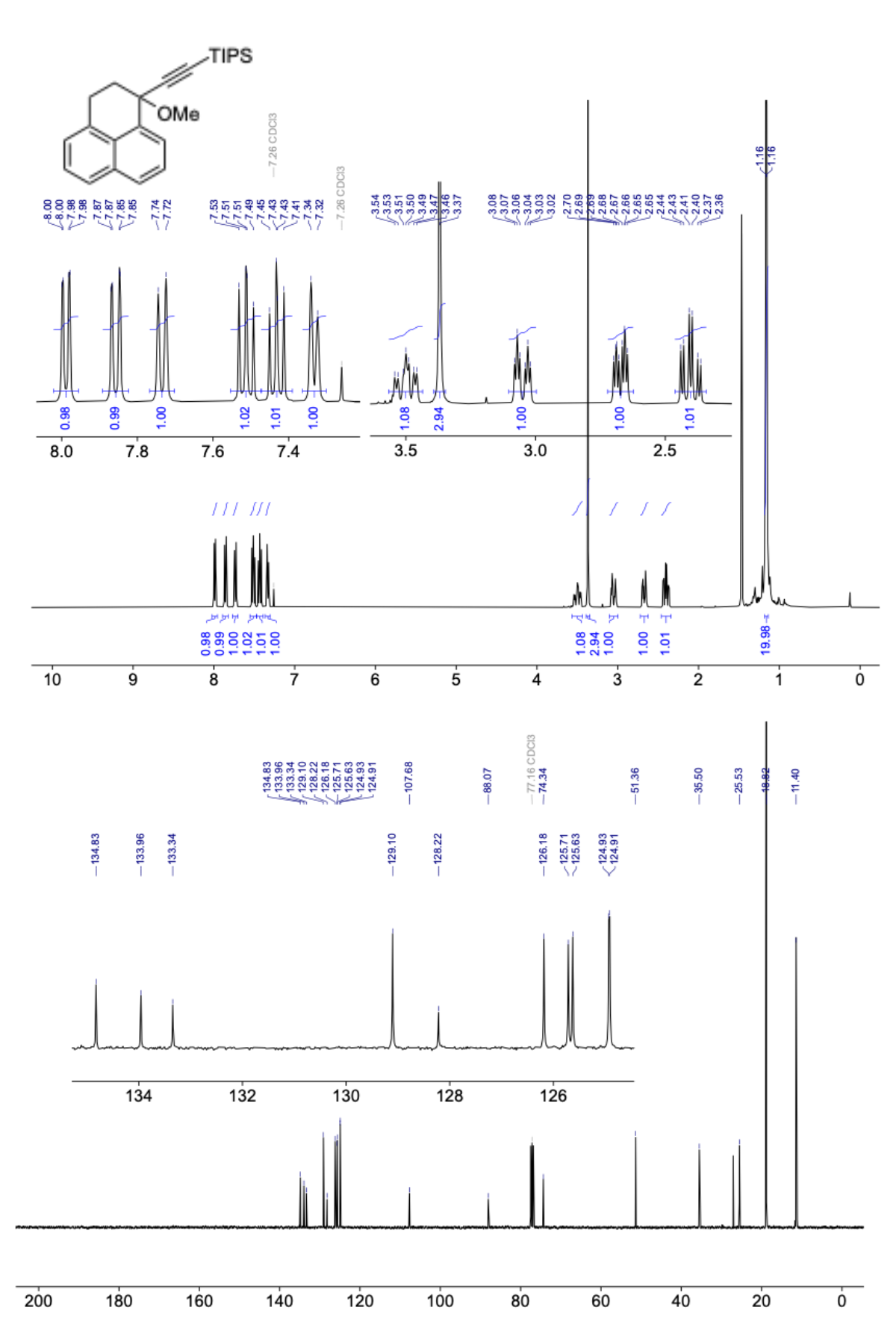}
    \caption{$^1$H (top) and $^{13}$C NMR (bottom) copies for compound \textbf{3}, CDCl$_3$, 400 MHz.}
    \label{fig:enter-label}
\end{figure}

\begin{figure}[h!]
    \centering
    \includegraphics[width=.7\linewidth]{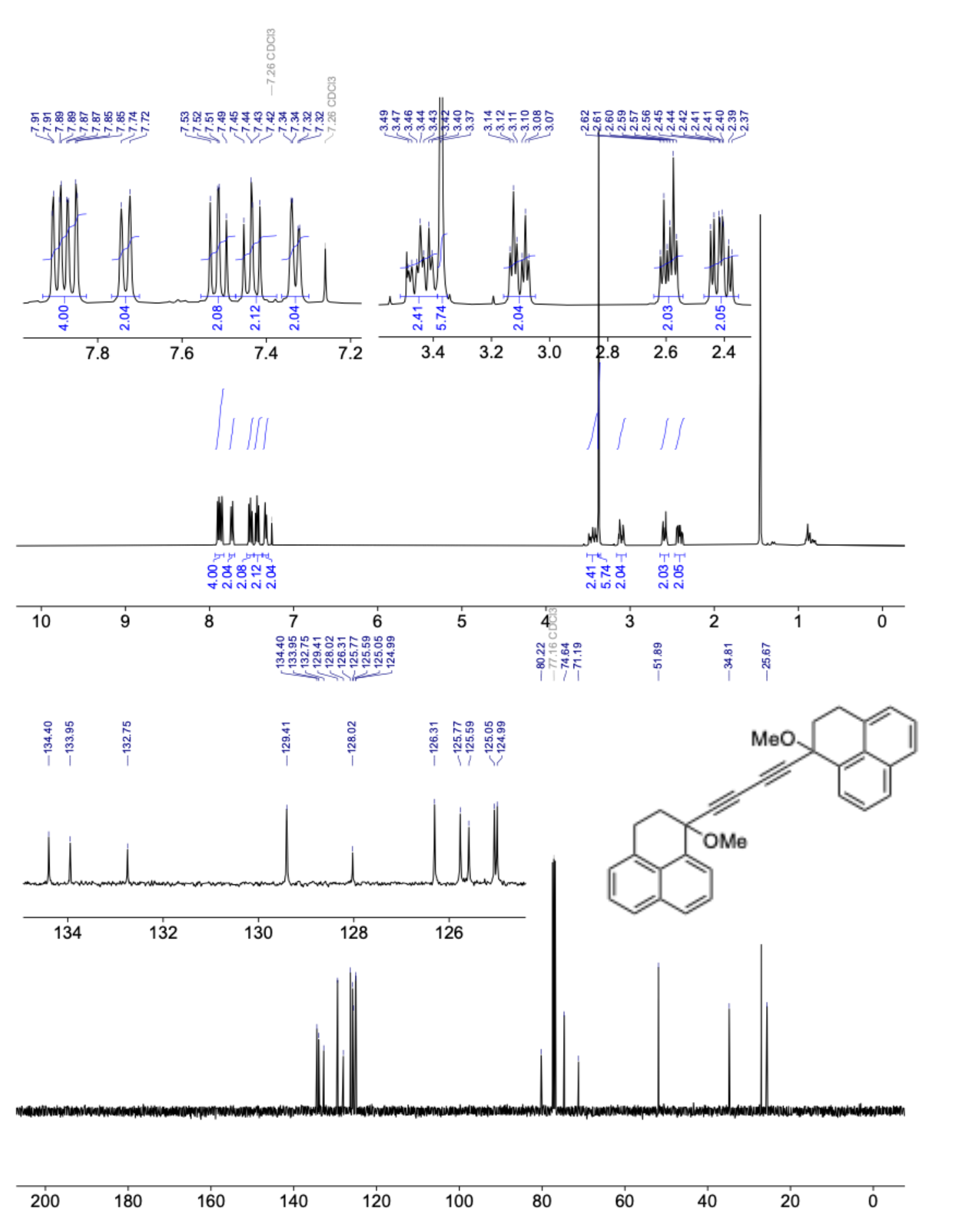}
    \caption{$^1$H (top) and $^{13}$C NMR (bottom) copies for compound \textbf{4}, CDCl$_3$, 400 MHz.}
    \label{fig:enter-label}
\end{figure}

\begin{figure}[h!]
    \centering
    \includegraphics[width=.8\linewidth]{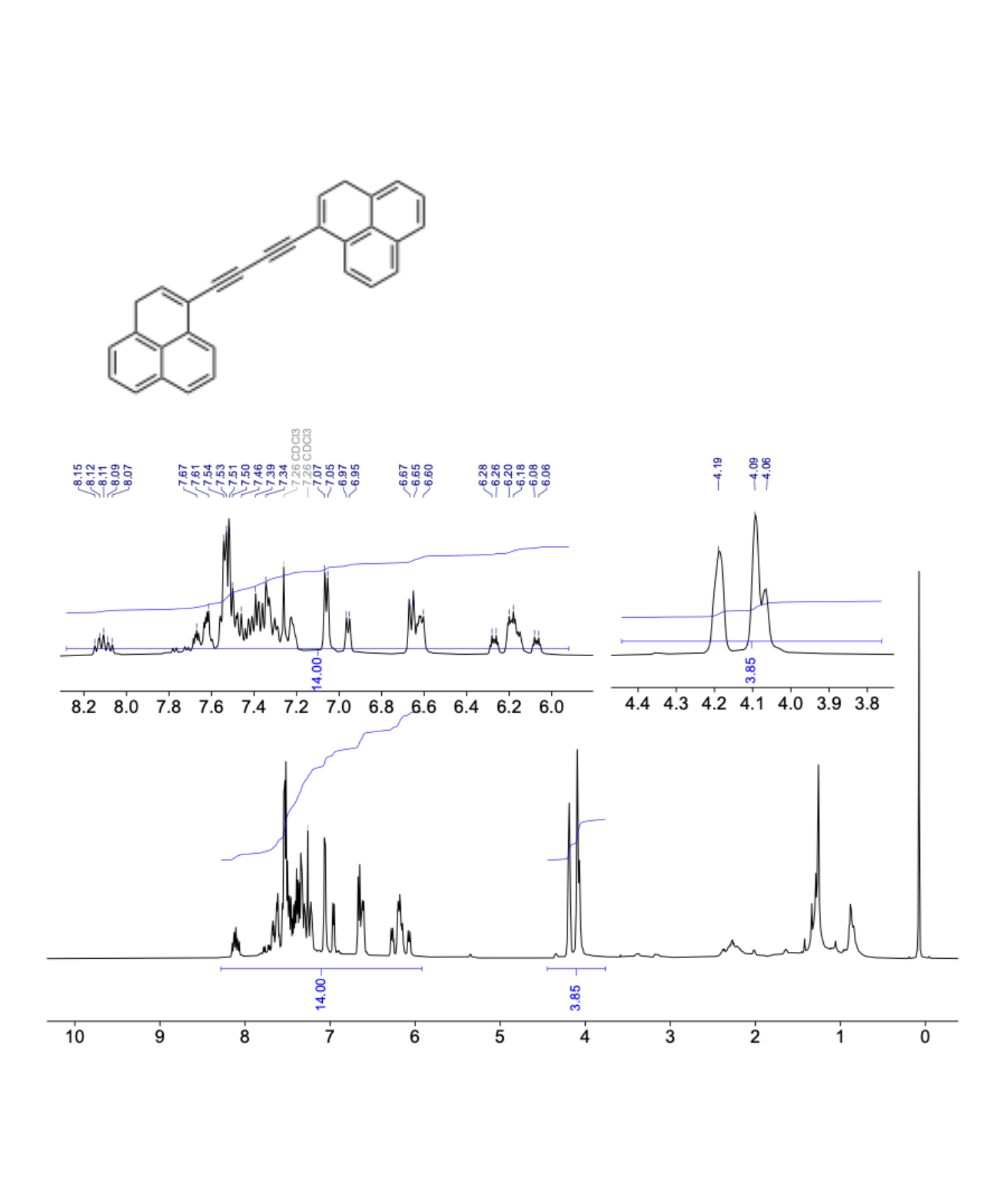}
    \caption{$^1$H NMR copy for compound \textbf{1p}, CDCl$_3$, 400 MHz.}
    \label{fig:enter-label}
\end{figure}

\clearpage
\noindent
\textbf{Copies of mass spectra:}

\begin{figure}[h!]
    \centering
    \includegraphics[width=.8\linewidth]{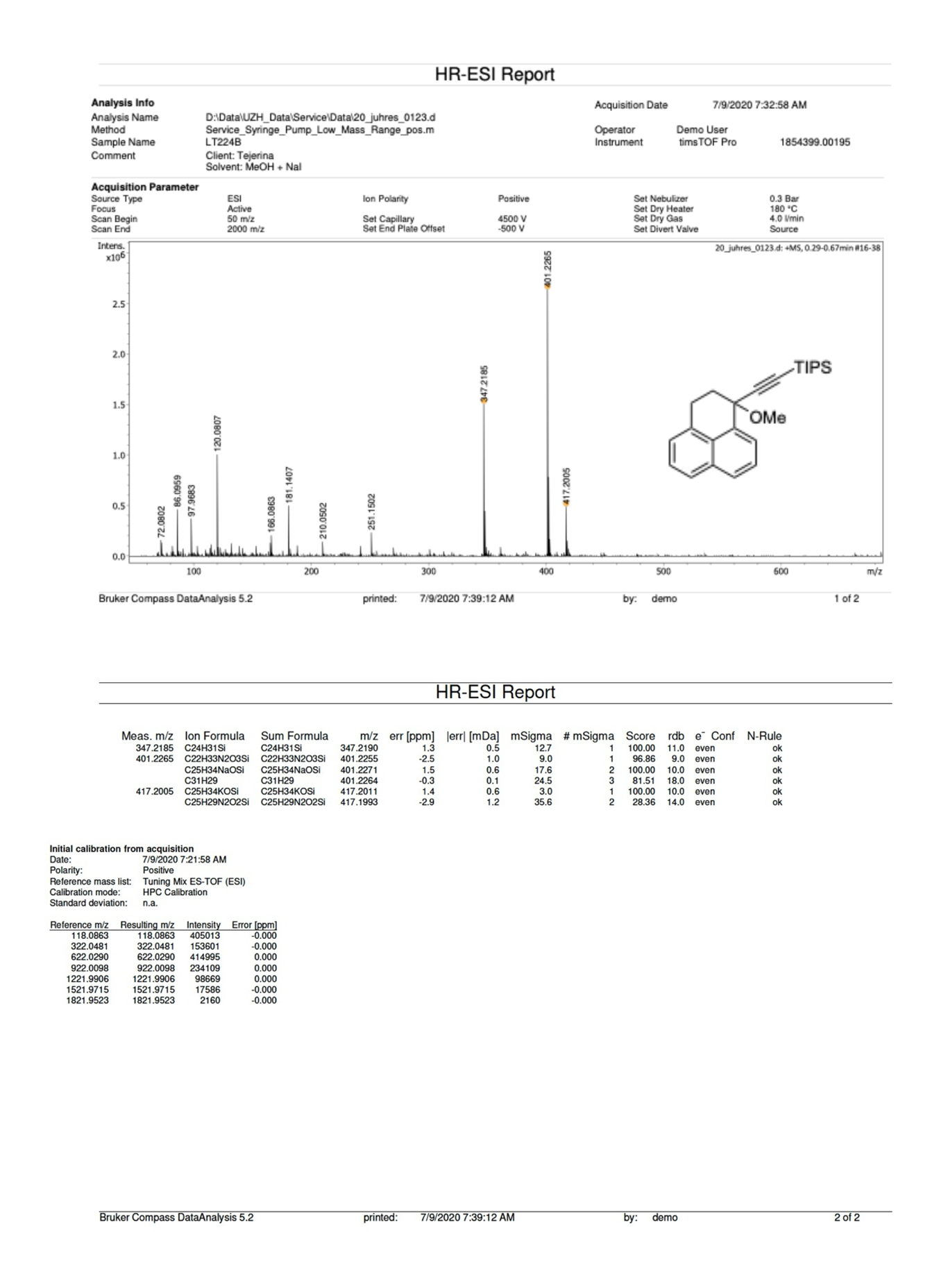}
    \caption{HRMS (ESI) copy for compound \textbf{3}.}
    \label{fig:enter-label}
\end{figure}

\begin{figure}[h!]
    \centering
    \includegraphics[width=.8\linewidth]{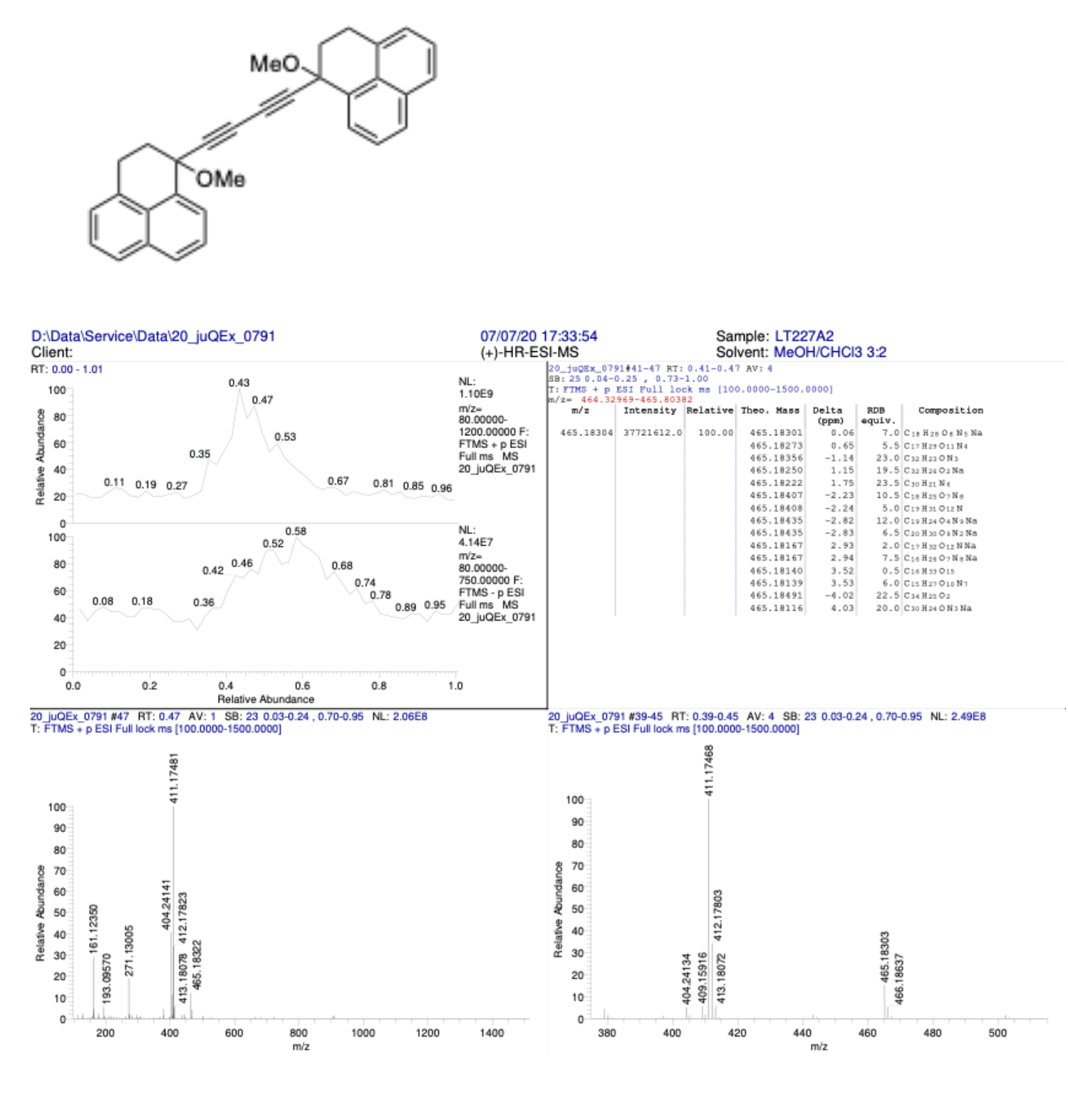}
    \caption{HRMS (ESI) copy for compound \textbf{4}.}
    \label{fig:enter-label}
\end{figure}

\clearpage

\section{Methods: sample preparation and scanning probe experiments}
STM and AFM measurements were conducted in a custom-built system operating under ultra-high vacuum (base pressure below $10^{-10}$ mbar) and at a temperature of 5 K. AFM measurements were performed with carbon monoxide (CO)-functionalized tips, and were performed in non-contact mode using a qPlus sensor\cite{giessibl_high-speed_1998} in frequency-modulation mode\cite{albrecht_frequency_1991} with a 0.5 Å oscillation amplitude. STM measurements were performed with metallic or CO-functionalized tips. The STM data were acquired in both constant-current and constant-height modes, while the AFM data were acquired in constant-height mode with $V = 0$ V. Positive (negative) values of the tip height $\Delta z$ indicate retraction (approach) from the STM set-point. The $dI/dV(V)$ spectra were  obtained by numerical differentiation of the corresponding $I(V)$ spectra. STM and AFM images, and spectroscopy curves, were post-processed using Gaussian low-pass filters.

\begin{figure}
    \centering
    \includegraphics[width=\linewidth]{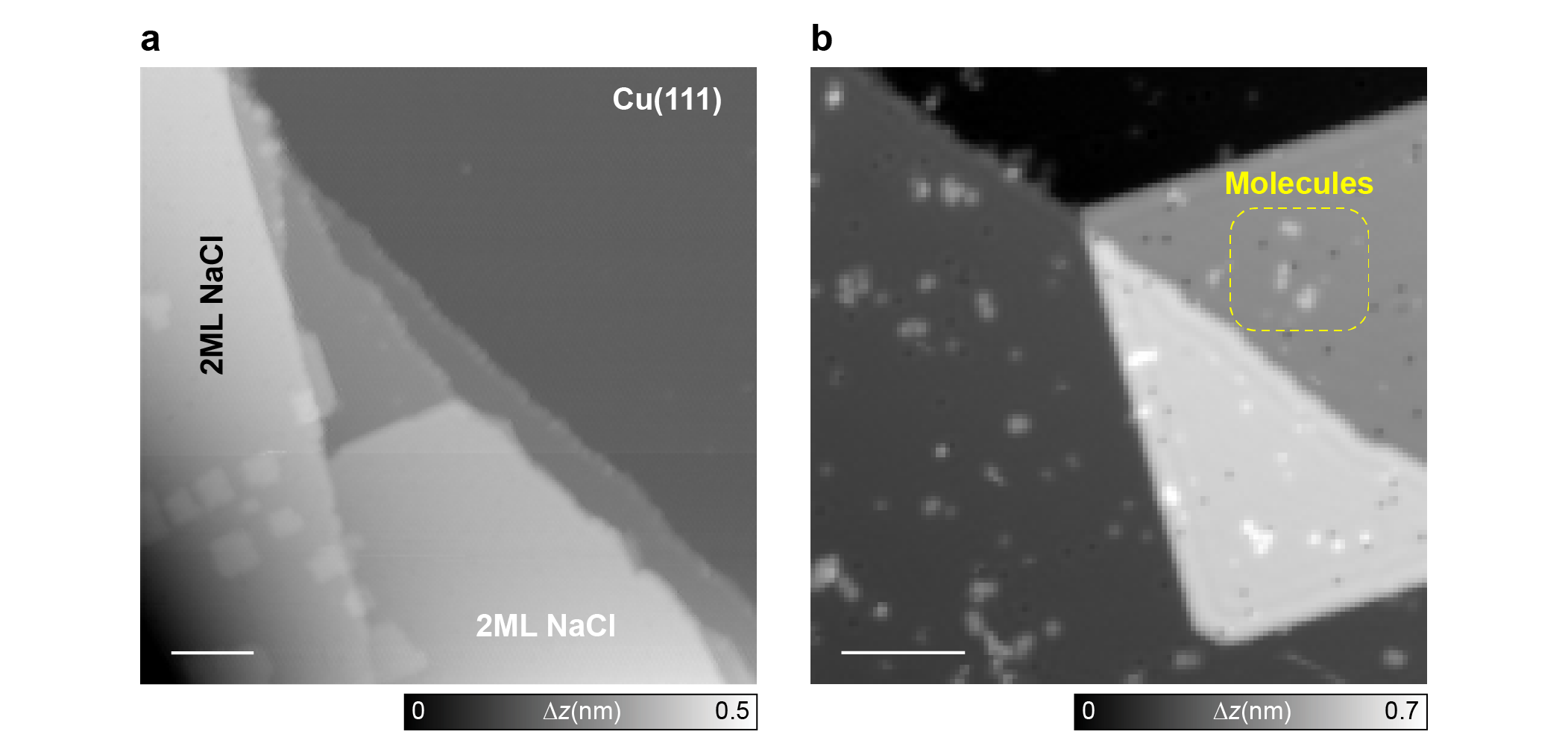}
    \caption{Overview STM images. (a) STM image of the NaCl/Cu(111) sample, showing predominantly bilayer (denoted 2ML) NaCl islands  along with few third-layer islands. (b) STM image showing a submonalyer coverage of \textbf{1p} on NaCl/Cu(111). The yellow dashed rectangle highlights three \textbf{1p} molecules adsorbed on a bilayer NaCl island. CO molecules are imaged as small circular depressions on NaCl and Cu(111) surfaces. STM set-points: $V$ = --0.7 V and $I$ = 4.7 pA (a); $V$ = 0.2 V and $I$ = 2.8 pA (b). The data were acquired with a metallic tip. Scale bars: 10\,nm.
    }
    \label{fig:STM overview}
\end{figure}

The Cu(111) surface was cleaned by repeated cycles of sputtering with Ne\textsuperscript{+} ions and annealing to 800 K. NaCl was thermally evaporated on Cu(111) held at 283 K, which led to predominantly bilayer (100)-terminated islands, with a minority of third-layer islands (Fig. \ref{fig:STM overview}a). Submonolayer coverage of \textbf{1p} was achieved by flashing an oxidized silicon wafer with the molecules in front of the cold sample in the microscope (Fig. \ref{fig:STM overview}b). CO molecules for tip functionalization were dosed from the gas phase onto the cold surface.

To generate \textbf{1} from \textbf{1p}, the STM tip was positioned at the center of a phenalene unit at \textit{V} = 0.2 V and \textit{I} = 1.0 pA, and the feedback loop was opened. The tip was then retracted by $\geq 5$ ~\AA{} (to limit the tunneling current) and the voltage was increased to 4--5 V. This protocol resulted in the homolytic cleavage of the C(\textit{sp}\textsuperscript{3})--H bond at each phenalene unit, leading to the generation of \textbf{1} via the singly dehydrogenated intermediate \textbf{1$'$} (Fig.~\ref{fig:manipulation}).

\section{Methods: theory}

\subsection{Ab-initio calculations}

The geometry of \textbf{1} (Fig.~\ref{fig:dft_cumulene}) was optimized using unrestricted Kohn–Sham DFT at the PBE0/def2-TZVP level, as implemented in the ORCA 6.0.1 software package, without imposed symmetry constraints\cite{orca,orca_5}. The optimized geometry revealed three distinct bond lengths in the central part of the molecule: 1.314~Å \;(bond $b$\textsubscript{A}), 1.241~Å \;(bond $b$\textsubscript{B}) and 1.358~Å \;(bond $b$\textsubscript{C}). The central bond $b$\textsubscript{A} is shorter than the terminal bond $b$\textsubscript{C}, while the bond $b$\textsubscript{B} is the shortest. To contextualize these bond lengths in terms of typical \ch{C\bond{sb}C} single, double, and triple bonds at the PBE0 level of theory, we also optimized ethylene and propyne molecules. Ethylene exhibits a \ch{C\bond{sb}C} bond length of 1.323~Å corresponding to a double bond, while propyne exhibits \ch{C\bond{sb}C} bond lengths of 1.450~Å and 1.200~Å, corresponding to prototypical single and triple bonds, respectively.

To assess the impact of bond length on the diradical index ($\gamma$), we constructed three models in which the \ch{C\bond{sb}C} bond lengths were fixed to that of standard \ch{C\bond{sb}C} single, double, or triple bond values:

    \textbf{Model 1}, in which the bond lengths of $b$\textsubscript{A}--$b$\textsubscript{C} were set equal to the \ch{C\bond{sb}C} bond length of ethylene (cumulenic geometry),
    
    \textbf{Model 2}, where the bond lengths of $b$\textsubscript{A} and $b$\textsubscript{C} were set to the bond length of the \ch{C\bond{sb}C} single bond in propyne, while the bond length of $b$\textsubscript{B} was set to that of the \ch{C\bond{sb}C} triple bond in propyne (polyynic geometry), and
    
    \textbf{Model 3}, which retained the bond lengths obtained from the PBE0 geometry optimization (unconstrained geometry).

These models yield bond length alternation (BLA) (defined as $b$\textsubscript{A} -- $b$\textsubscript{B}) of 0.000~Å, 0.250~Å, and 0.073~Å, respectively.

State-specific complete active space self-consistent field (CASSCF) calculations were performed in a CAS(12,12)\cite{RN155} using the optimized geometries from all three models, thereby expanding the electronic wavefunction into all possible configuration state functions to determine the neutral singlet ground state. To accelerate the SCF procedure, the RIJCOSX approximation was employed, along with the def2/JK auxiliary basis set for exchange fitting\cite{orca_def_jk}. Initial orbitals were generated from quasi-restricted natural orbitals obtained at the PBE0 level. 

Following the definition of the minimal multiconfigurational representation of the wavefunction provided in the main manuscript (Eq.~1), we extracted the coefficients $\mathrm{A_1}$ and $\mathrm{A_2}$ corresponding to two spin determinants\cite{RN230}, namely, the bonding ($ \mathrm{A_1}|\,\mathrm{222222000000}\,\rangle$) and the antibonding ($ \mathrm{A_2}|\,\mathrm{222220200000}\,\rangle$) configurations.
In these kets, the sixth and seventh entries correspond to CASSCF natural orbitals that resemble the HOMO and LUMO (Fig. \ref{fig:no_casscf}), respectively, and each entry’s value denotes the electron occupation of the corresponding orbital.

\vspace{-0.5cm}
\begin{equation*}
\begin{aligned}
\mathrm{Model}\ \mathrm{1}\ \mathrm{(cumulenic)}  &= -0.880432257\,|\,\mathrm{222222000000}\,\rangle + 0.234255081\,|\,\mathrm{222220200000}\,\rangle, \\
\mathrm{Model}\ \mathrm{2}\ \mathrm{(polyynic)} &= -0.706808652\,|\,\mathrm{222222000000}\,\rangle + 0.526172477\,|\,\mathrm{222220200000}\,\rangle, \\
\mathrm{Model}\ \mathrm{3}\ \mathrm{(unconstrained)} &= -0.851882960\,|\,\mathrm{222222000000}\,\rangle + 0.339288678\,|\,\mathrm{222220200000}\,\rangle.
\end{aligned}
\end{equation*} 

These coefficients were used to compute $\gamma = 2|\mathrm{A_2}|^2$ for each model (Table \ref{tab:casscf_summary}). Since our active space is CAS(12,12), rather than the minimal CAS(2,2) representation, we introduced a normalized quantity, $\gamma_{\mathrm{norm}}$, which accounts for the total weight of both configurations. Specifically, we used the expression:

\begin{equation}
\gamma_{norm} = \frac{\gamma}{|\mathrm{A_1}|^2 + |\mathrm{A_2}|^2 }
\end{equation}

The excited states energies for both the neutral and charged states were determined via state-averaged CASSCF (SA-CASSCF) calculations. The resulting energies were corrected for dynamic correlation using the domain-based local pair natural orbital N-electron valence perturbation theory to second order (DLPNO-NEVPT2)\cite{dlpno_nevpt2}. For the SA-CASSCF calculations, the neutral states were averaged over 8 states for both the triplet and singlet multiplicities. For the cationic and anionic states of the molecule, CAS(11,12) and CAS(13,12) were used, respectively, with 8 doublet excitations considered for each charge state. Table \ref{tab:casscf_summary} presents a summary of the results from these calculations.

\renewcommand{\arraystretch}{1.2}
\begin{table}
    \centering
    \begin{tabular}{cccccccc}
    \hline
    System &$\gamma$ & $\gamma_{norm}$ & $\mathrm{S_0} \to \mathrm{S_1}$ [eV]  &  $\mathrm{S_0} \to \mathrm{T_1}$ [eV] &  $\mathrm{D_0^+} \to \mathrm{D_1^+}$ [eV] & $\mathrm{D_0^-} \to \mathrm{D_1^-}$ [eV] \\
    \hline
    Model 1  & 0.110 & 0.132 & 1.637 & 0.734 &1.103 & 1.171 \\
    (cumul.)  & & & & & & \\
     & & & & & & \\
    Model 2  & 0.554 & 0.713 & 1.831 & 0.108 &0.439 & 0.414\\
    (polyyn.)  & & & & & & \\
     & & & & & & \\
    Model 3  & 0.230 & 0.274 & 1.575 & 0.427 &0.677  & 0.614  \\
    (uncons.)  & & & & & & \\
    \hline
    \end{tabular}
    \caption{Summary of the CASSCF results for the diradical character $\gamma$ and excitation energies of the neutral and charged states.}
    \label{tab:casscf_summary}
\end{table}

\subsection{Model Hamiltonian and DMRG calculations}

\textbf{TB model and validation by DFT.} To model \textbf{1}, we started by defining an effective tight-binding (TB) Hamiltonian.
Although the C$_4$ chain features $sp$ hybridization, we considered only the $p_z$ orbital of each of the sp-hybridized carbon atoms.
The reasoning behind this approximation is that, in a planar configuration, the $p_z$ orbitals of the phenalenyl units only couple to the $p_z$ orbitals of the chain, being orthogonal to the in-plane orbitals of the chain.
DFT calculations confirmed that the remaining four electrons in the four in-plane orbitals of the chain hybridize and form orbitals far away ($\gtrsim 2$~eV) from the Fermi energy. We thus disregard these in-plane orbitals in our model.
As for the hoppings, we considered only nearest-neighbor terms, as the phenalenyl units are connected via their majority sublattice sites, where their unpaired electron wavefunction resides.
In the phenalenyl units, we took the standard value for nanographenes, $t_1 = -2.7$~eV.
In the carbon chain, we parameterized the hoppings according to the C-C distances, which we obtained from the three geometry optimizations described in the section above.
In particular, we used the parameterization from Ref.~\citenum{catarina_conformational_2024}, obtaining: (i) for the cumulenic case (where all bonds in the chain have C-C bond lengths equal to those of the double bonds obtained from ethylene), \textit{t}\textsubscript{chain,\textit{b}\textsubscript{A}} = \textit{t}\textsubscript{chain,\textit{b}\textsubscript{B}} = \textit{t}\textsubscript{chain,\textit{b}\textsubscript{C}} = --3.0 eV; (ii) for the polyynic case (where we assumed alternating single and triple bonds, with C-C bond lengths obtained from propyne), \textit{t}\textsubscript{chain,\textit{b}\textsubscript{A}} = \textit{t}\textsubscript{chain,\textit{b}\textsubscript{C}} = --2.6 eV, \textit{t}\textsubscript{chain,\textit{b}\textsubscript{B}} = --3.5 eV; (iii) for the unconstrained geometry optimization of the system, \textit{t}\textsubscript{chain,\textit{b}\textsubscript{A}} = --3.1 eV, \textit{t}\textsubscript{chain,\textit{b}\textsubscript{B}} = --3.3 eV, \textit{t}\textsubscript{chain,\textit{b}\textsubscript{C}} = --2.9 eV.
This parameterization was validated by comparing the TB energy levels and orbitals with DFT calculations, which yielded good agreement close to the Fermi energy (Fig.~\ref{fig:TB_DFT_validation}).
DFT calculations were carried out with Quantum Espresso\cite{giannozzi_quantum_2009}, using the Perdew-Burke-Ernzerhof functional\cite{perdew_generalized_1996}.
We used the pseudopotentials recommended for precision in the SSSP library\cite{prandini_precision_2018} (version 1.3.0), together with the suggested kinetic energy cutoffs of 80~Ry for the wavefunction and 360~Ry for the charge density.
To simulate the molecule, we considered a 14~{\AA} vacuum separation between unit cells in all directions.
For that reason, our grid of reciprocal points included only the $\Gamma$-point.
We performed spin-restricted self-consistent field calculations until energies were converged up to $8 \times 10^{-5}$~Ry, starting from the three optimized geometries described in section III.1.
\vspace{0.5cm}

\textbf{Modeling interactions and DMRG details.} Previous works on phenalenyl-based magnetic nanostructures~\cite{jacob_theory_2022,krane_exchange_2023,catarina_conformational_2024} have shown that modeling interactions in this type of systems is difficult.
On the one hand, methods based on the CAS approximation have convergence issues.
For this reason, here we relied on density matrix renormalization group (DMRG)~\cite{white_density_1992} calculations.
On the other hand, the parameterization of the interaction parameters, especially their long-range behavior and dependence on the underlying surface, remains largely unknown.
In light of this, here we adopted a simplified approach and modeled interactions via a Hubbard term with on-site Hubbard repulsion $U = 2|t_1|$.
The resulting TB-Hubbard Hamiltonian was solved using DMRG as implemented in the ITensor library~\cite{fishman_itensor_2022}.
We performed DMRG sweeps with an adaptive scheme where the maximum bond dimensions were allowed to grow indefinitely in order to maintain the truncation error below $10^{-6}$, which ensures high accuracy.

Using DMRG for the system with unconstrained geometry, we obtained a singlet-triplet ($S_0 \rightarrow T_1$) excitation energy of 361~meV, in reasonable agreement with the CASSCF calculations including perturbative corrections for dynamic correlations.
Moreover, in the $-1$ charged sector, we obtained a doublet-doublet ($D_0^- \rightarrow D_1^-$) excitation energy of 568~meV, in good agreement not only with the DLPNO-NEVPT2-corrected CASSCF calculations but also with the experimental measurements.
\vspace{0.5cm}

\textbf{Natural orbital analysis.} Given a many-body state $| \psi \rangle$ obtained from DMRG, its natural orbitals and the corresponding occupation numbers were computed by diagonalizing the one-electron reduced density matrix,
\begin{equation}
    \rho^{(1)}_{i,i'} = \langle \psi | \sum_{\sigma} \hat{c}^\dagger_{i,\sigma} \hat{c}_{i',\sigma} | \psi \rangle,
\end{equation}
where $\hat{c}_{i,\sigma}$ ($\hat{c}^\dagger_{i,\sigma}$) denotes the annihilation (creation) operators for an electron in site $i$ with spin $\sigma=$ {$\uparrow,\downarrow$}.
In Fig.~\ref{fig:NO_analysis}, we show the natural orbital analysis obtained for the ground state of the system with unconstrained geometry, found to be a singlet.
These results corroborate the picture described in Eq. 1 of the main text, as only two natural orbitals have occupancies significantly different from 0 and 2, and their shapes resemble those of the HOMO and LUMO.

As an estimate of the number of unpaired electrons, we employed the expression 
\begin{equation}
    N_u = \sum_\lambda n_\lambda^2 (2 - n_\lambda)^2,
\end{equation}
as outlined in Ref.~\citenum{head-gordon_characterizing_2003}, where $n_\lambda$ are the occupation numbers of the natural orbitals.
The ground-state results presented in Table~\ref{tab:A1_A2} reveal the expected trend of increasing radical character from the cumulenic to the polyynic geometry.

\begin{table}
    \centering
    \begin{tabular}{c| c c c c c }
        & $N_u$ & $A_1$ & $A_2$ & $\gamma=2|A_2|^2$ & $\text{norm} = |A_1|^2 + |A_2|^2$\\
        \hline
        Model 1 (cumulenic) & 0.60 & $-0.79$ & 0.19 & 0.07 & 0.65 \\
        Model 2 (polyynic) & 1.26 & $-0.73$ & 0.35 & 0.25 & 0.66 \\
        Model 3 (unconstrained) & 0.71 & 0.78 & $-0.24$ & 0.11 & 0.66 \\
    \end{tabular}
    \caption{Number of unpaired electrons $N_u$, multiconfigurational coefficients $A_1$ and $A_2$ defined in Eq.~(1) of the main text, diradical index $\gamma$, and norm of the associated two-configuration picture, all computed by DMRG for the ground state of the system with three different geometries.}
    \label{tab:A1_A2}
\end{table}

To obtain the coefficients $A_1$ and $A_2$ defined in Eq.~1 in the main text, we followed Ref.~\citenum{catarina_conformational_2024} and computed the overlap between the ground state obtained by DMRG and many-body states created by starting with a vacuum state and applying creation operators to fill the targeted natural orbitals.
Results obtained for the three geometries are summarized in Table~\ref{tab:A1_A2}.
Analogous overlap calculations were carried out to validate the scheme shown in Fig.~3b of the main text.
\vspace{0.5cm}

\textbf{Dyson orbitals.} Dyson orbitals characterize transitions between states whose charge differs by $\pm1$.
They are particularly relevant to describe elastic excitations probed by scanning tunneling microscopy/spectroscopy (STM/STS) in multiconfigurational systems\cite{KumarJACS2025}. 
In this context, and disregarding population effects (e.g., due to temperature or bias), let us consider transitions between a neutral ground state with, say, $N$ electrons, denoted by $|\psi_0^N\rangle$, to the $m$-th charged state with $N\pm1$ electrons, denoted by $| \psi_m^{N\pm1} \rangle$.
Such an excitation from $|\psi_0^N\rangle$ to $| \psi_m^{N\pm1} \rangle$ occurs at positive/negative bias, and corresponds to a negative/positive ion resonance (NIR/PIR), which we label as NIR$+m$/PIR$-m$.
The corresponding spin-resolved ($\sigma=$ $\uparrow,\downarrow$) Dyson orbitals are written as $\mathcal{D}_{\text{NIR}+m}^{\sigma} (\bm{r})$ and $\mathcal{D}_{\text{PIR}-m}^{\sigma} (\bm{r})$, where $\bm{r} = (x,y,z)$ denotes the spatial coordinates.

Since our model Hamiltonian preserves spin rotational invariance, the local density of the Dyson orbitals is independent of the spin, i.e., $|\mathcal{D}_{\text{NIR}+m/\text{PIR}-m}^{\sigma=\uparrow} (\bm{r})|^2$ = $|\mathcal{D}_{\text{NIR}+m/\text{PIR}-m}^{\sigma=\downarrow} (\bm{r})|^2$.
Within our model framework, based on a carbon $2p_z$ atomic orbital basis and solved via DMRG, the Dyson orbitals can be expressed as
\begin{equation}
    \mathcal{D}_{\text{NIR}+m}^{\sigma} (\bm{r}) = \sum_i \mathcal{D}_{\text{NIR}+m}^{\sigma} (i) \phi_{i} (\bm{r}),
\end{equation}
\begin{equation}
    \mathcal{D}_{\text{PIR}-m}^{\sigma} (\bm{r}) = \sum_i \mathcal{D}_{\text{PIR}-m}^{\sigma} (i) \phi_{i} (\bm{r}),
\end{equation}
where 
\begin{equation}
    \phi_{i} (\bm{r}) = (z-z_i)\mathrm{e}^{-|\bm{r}-\bm{r}_i|/r_0},
\end{equation}
with $r_0 = 0.325$~{\AA}, denotes the $2p_z$ orbital of a carbon atom at site $i$, with position $\bm{r}_i = (x_i,y_i,z_i)$, and
\begin{equation}
    \mathcal{D}_{\text{NIR}+m}^{\sigma} (i) = \langle \psi_m^{N+1} | \hat{c}^\dagger_{\sigma}(i) | \psi_0^N \rangle,
\end{equation}
\begin{equation}
    \mathcal{D}_{\text{PIR}-m}^{\sigma} (i) = \langle \psi_m^{N-1} | \hat{c}_{\sigma}(i) | \psi_0^N \rangle.
\end{equation}
When accounting for population effects, for example, through a rate equation formalism (as described below), it may also be necessary to compute Dyson orbitals from excited states of the neutral system, for which the calculations are straightforward.
In Fig.~\ref{fig:Dyson transitions}, we show the local density of the relevant Dyson orbitals for our system.
\vspace{0.5cm}

\textbf{Rate equations.} To reproduce the experimental results, we employed a rate-equation model using a steady-state master equation
\begin{equation}
    0 = \sum_{i\neq j} \left( N_i\Gamma_{i \rightarrow j} - N_j\Gamma_{j \rightarrow i} \right)
\end{equation}
with $\Gamma_{i \rightarrow j}$ being the transition rate from state $i$ to state $j$ and $N_i$ being the occupation probability of state $i$.
We consider six states, as depicted in Fig. 3b in the main text and listed in Table~\ref{tab:states}. Their energies were chosen such that the calculated d$I$/d$V$($V$) spectrum (Fig. \ref{fig:ME_spectroscopy}) matches the experimental one.
\begin{table}[]
    \centering
    \begin{tabular}{c c c}
        \textbf{State} & \textbf{Energy} & \textbf{Charge}\\
        \hline
        $S_0$ & 0.00\,eV & 0 \\
        $T_1$ & 0.30\,eV & 0 \\
        $D_0^-$ & 0.60\,eV & +1$e^-$ \\
        $D_1^-$ & 1.30\,eV & +1$e^-$ \\
        $D_0^+$ & 1.65\,eV & -1$e^-$ \\
        $D_1^+$ & 2.50\,eV & -1$e^-$ \\
    \end{tabular}
    \caption{Electronic states considered in the master equation.}
    \label{tab:states}
\end{table}
    
For the transition rates between these states we assume three pathways: charge transfer between molecule and surface $\Gamma^S$, charge transfer between tip and molecule $\Gamma^T$, and a charge-neutral decay rate from $T_1$ state to $S_0$.
The coupling to the surface is modeled by a single tunneling barrier with height $\Phi_\mathrm{eff}$ and width $d=4$ \AA, whereas the tunnel rate at $d=0$ is defined to correspond to a quantum of conductance $2e/h$.
The exact values for $\Gamma^S$ are not crucial for the simulation, since the total current is limited by the coupling to the tip, i.e., $\Gamma^T \ll \Gamma^S$.

The coupling to the tip is based on the Dyson orbitals obtained by DMRG (see above) and the assumption of an s-wave tip. The low tunneling currents in experiment -- and consequently large tip-sample distances -- results in a predominant s-wave tunneling, even with a CO-functionalized tip\cite{paschke_distance_2025}. To include the effect of a bias modulated tunneling barrier $\Phi_\mathrm{eff}$, the Dyson orbitals were calculated with Slater-type $p_z$ orbitals at a constant-height plane $1.5$ \AA\ above the molecule and exponentially extrapolated into the vacuum. 
The effective tunneling barrier height is defined as $\Phi_\mathrm{eff}=\Phi+\Delta E_{fi}\Delta Q_{fi}+eV/2$, with workfunction $\Phi=4.9$ \,eV, bias voltage $V$ and $\Delta E_{fi}$ and $\Delta Q_{fi}$ being the difference in energy and charge, respectively, between final and initial state.
An empirical conversion factor $\eta$ is used to translate the square of the extrapolated wavefunction into a hopping rate, that matches the experiment.

The effect of high electron-phonon coupling in the underlying NaCl surface~\cite{Fatayer2018NatNano,Vasilev2024ACSNano,Sellies2025NatNano} was captured by assuming a shift of $E_r=0.2$\, eV to higher energies for all charging transitions ($\Delta Q_{fi} \neq 0$) and a gaussian broadening with half width at half maximum of $0.12$\, eV.

To account for the different multiplicities of the states, all transition rates were multiplied with an additional factor $\Upsilon_{fi}$, based on the Clebsch-Gordon coefficients. The transition from singlet to doublet being $\Upsilon_{DS} = 2$, from doublet to triplet $\Upsilon_{TD}=1.5$ and all other allowed transitions being $\Upsilon_{fi}=1$.
The master equation was solved for each respective tip position and voltage to obtain the steady-state occupation probabilities, and the tunneling current calculated from the net charge transfer between the molecule and surface. A good match with experiment was obtained for an empirical conversion factor of $\eta=6\times10^9$ and a triplet lifetime of 30\,ns.
Note that the triplet lifetime in the simulations is based on several assumptions, e.g., s-wave tip or the precise relative intensities of orbitals at a given tip height. However, our experiments do not provide a direct measurement of intrinsic triplet lifetimes, which would be accessible by other methods using thicker insulating films\cite{PengScience2021}. The 30\,ns used in our simulations can only be regarded as a rough estimate for the order of magnitude.
Figure~\ref{fig:ME_spectroscopy} shows representative calculated $I(V)$ and d$I$/d$V$($V$) spectra for a tip placed at $z=11$ \AA\ near the center of the molecule, comparable to the experimental spectrum shown in Fig. 3a of the main text. The calculated transition probability maps are displayed in Fig. \ref{fig: failure of single particle} for $V=-1.8$\,V, $z=10.0$\,\AA; $V=+0.9$\,V, $z=10.3$\,\AA; and $V=+1.5$\,V, $z=11.5$\,\AA.

\newpage
\section{Experimental and theoretical data}

\begin{figure}[h]
    \centering
    \includegraphics[width=\linewidth]{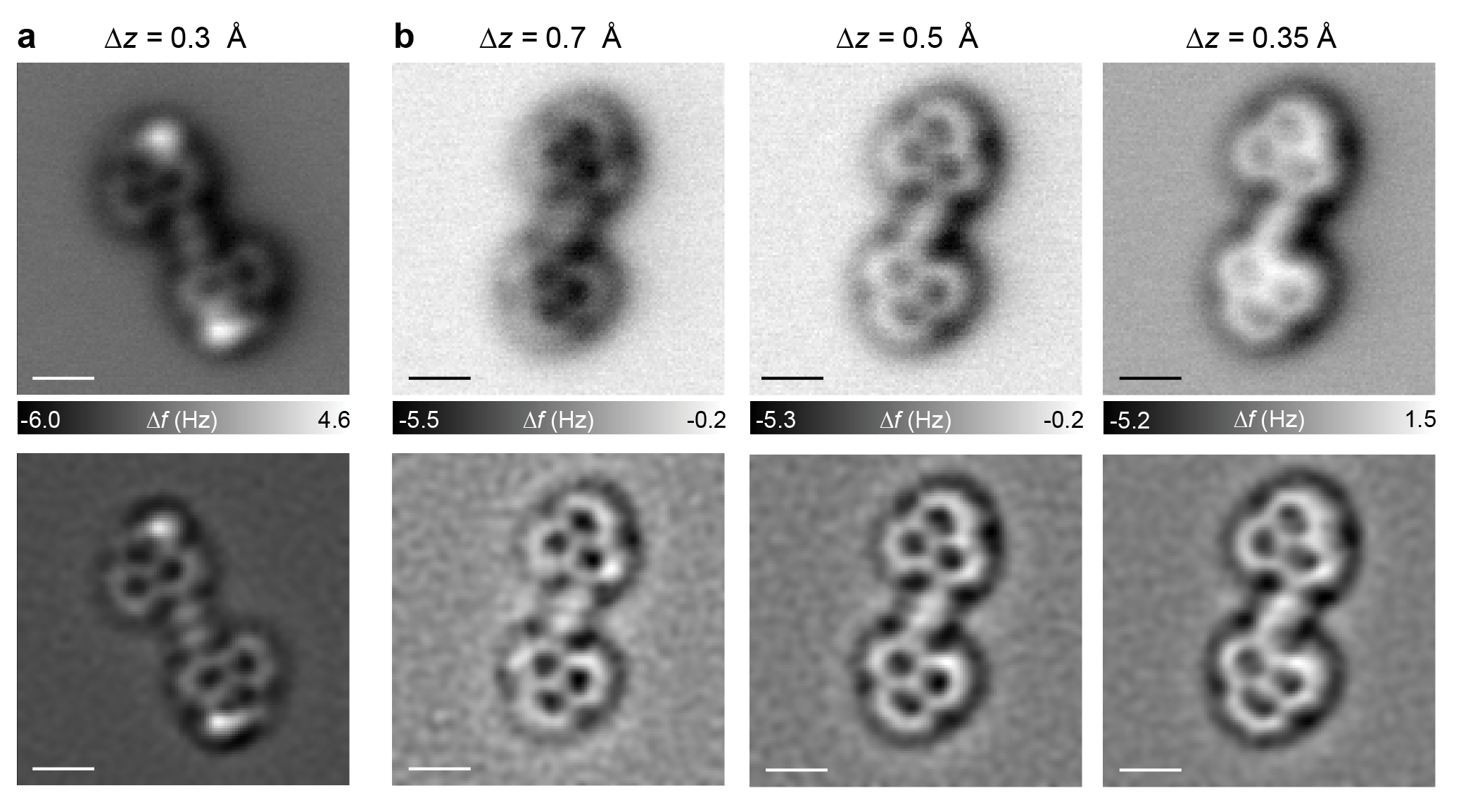}
    \caption{AFM images of \textit{trans} \textbf{1p} (a) and \textbf{1} (b). Top: AFM images; bottom: corresponding Laplace-filtered versions. Qualitatively, the bond-order contrast in the C\textsubscript{4} chain of the \textit{trans} isomer appears similar to that of the \textit{cis} isomer. STM set-point: $V$ = 0.2 V and $I$ = 1.0 pA on NaCl. Scale bars: 0.5 nm.}
    \label{fig:trans AFM}
\end{figure}

\begin{figure}[h]
    \centering
    \includegraphics[width=\linewidth]{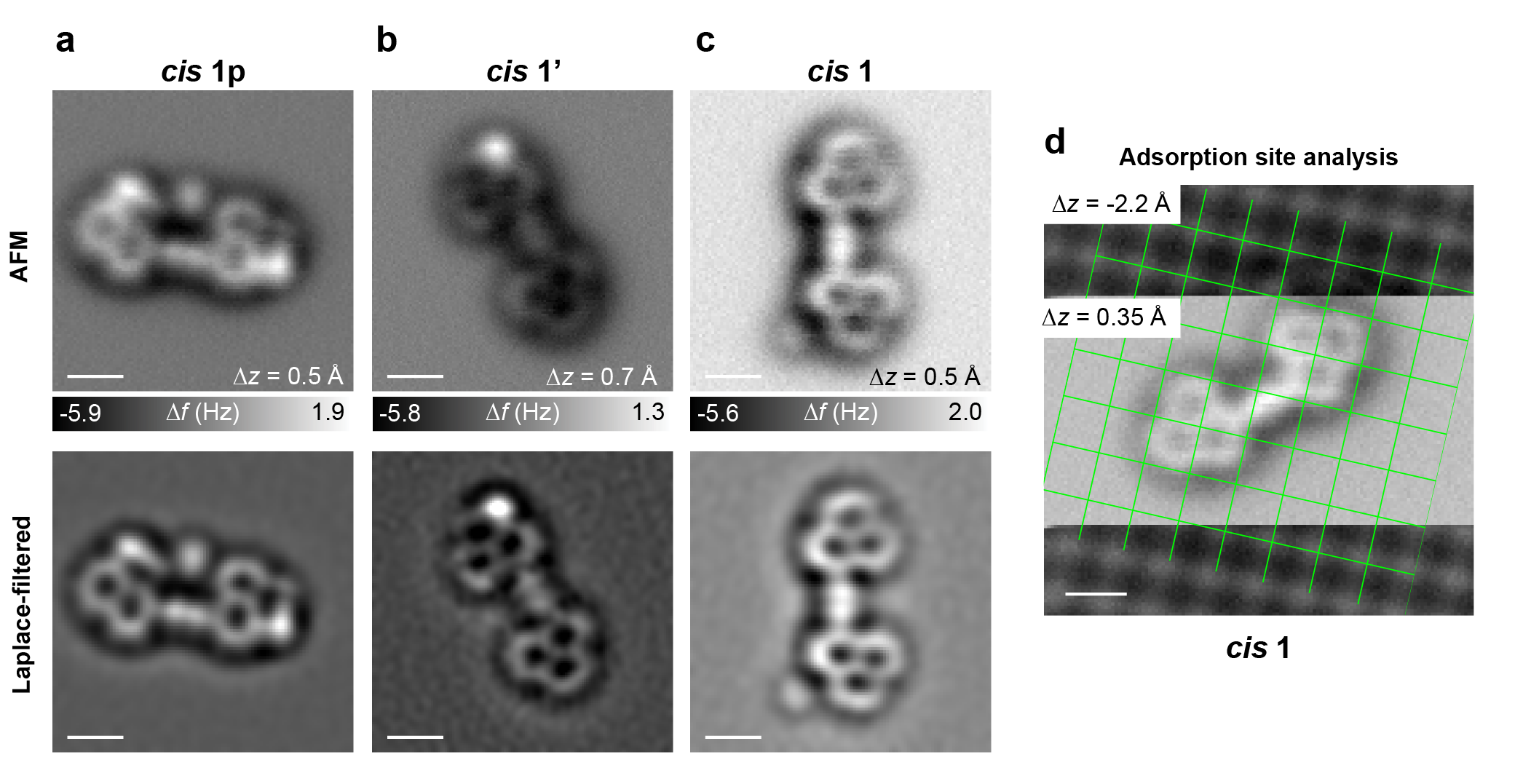}
    \caption{Stepwise dehydrogenation of \textit{cis} \textbf{1p}. (a--c) AFM images (top) with the corresponding Laplace-filtered images (bottom) of the precursor \textbf{1p} (a), the singly dehydrogenated intermediate \textbf{1$'$} (b) and the target molecule \textbf{1} (c). (d) AFM image where both the molecular structure of \textbf{1} and the NaCl lattice is resolved. Crossing points of the overlaid lattice (in green) correspond to the Cl\textsuperscript{--} sites of NaCl. STM set-point: $V$ = 0.2 V and $I$ = 1.0 pA on NaCl. Scale bars: 0.5 nm.}
    \label{fig:manipulation}
\end{figure}

\begin{figure}[h]
    \centering
    \includegraphics[width=\linewidth]{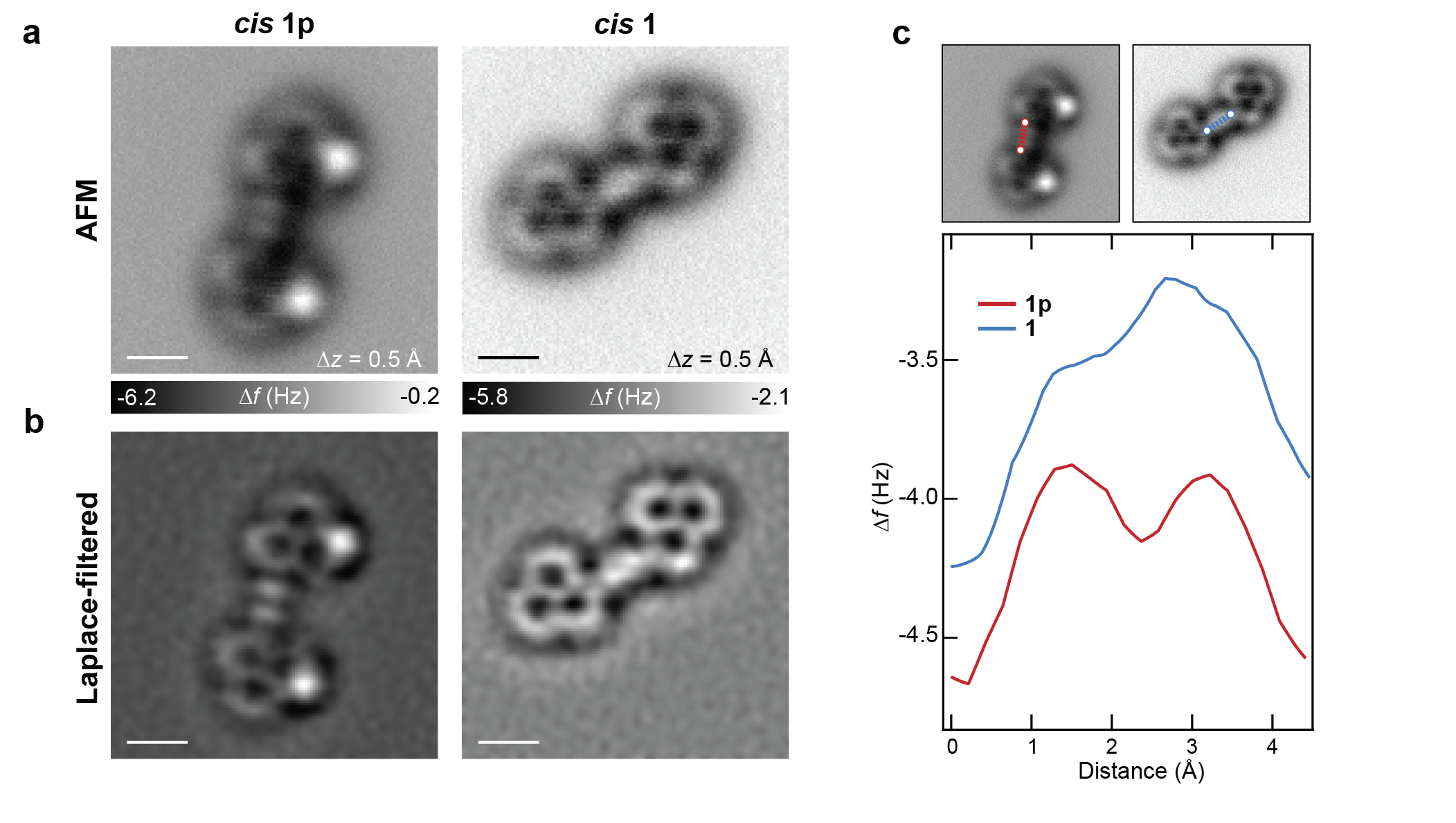}
    \caption{Bond-order contrast comparison with the same tip. (a, b) AFM images (top) and corresponding Laplace-filtered images (bottom) of \textit{cis} \textbf{1p} (a) and \textit{cis} \textbf{1} (b). (c) $\Delta f$ line profiles along the C\textsubscript{4} chains of \textbf{1p} and \textbf{1}. STM set-point: $V$ = 0.2 V and $I$ = 1.0 pA on NaCl. Scale bars: 0.5 nm.}
    \label{fig:same mol same tip}
\end{figure}

\begin{figure}[t]
  \centering
  \includegraphics[width=\columnwidth]{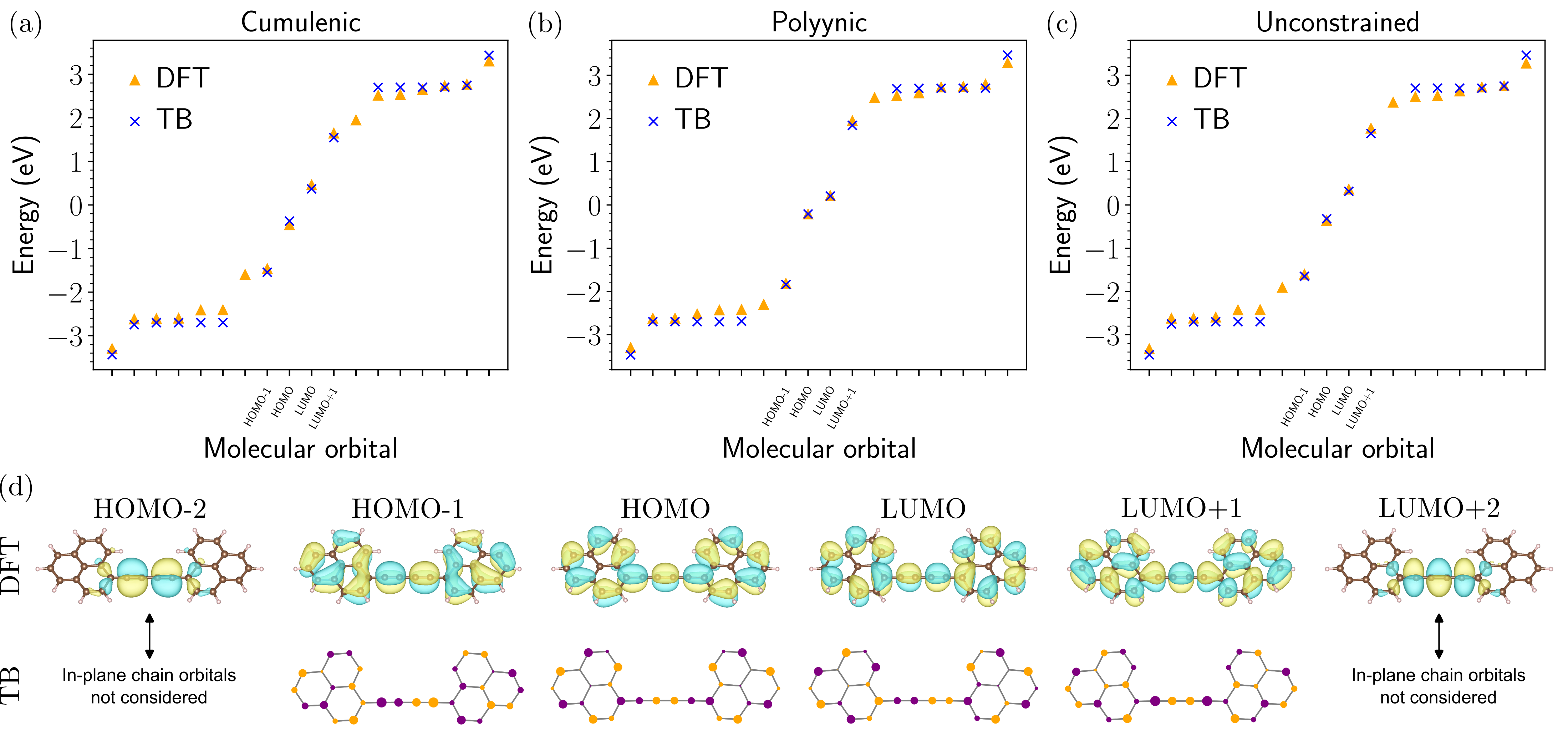}
  \caption{Validation of TB model by DFT. 
  (a--c) DFT and TB energy levels, obtained for (a) cumulenic, (b) polyynic, and (c) unconstrained geometries. States above and below zero energy are unoccupied and occupied, respectively. TB energy levels were shifted (laterally) at the HOMO-2 and LUMO+2 positions, as the corresponding DFT orbitals were found to be predominantly composed of in-plane orbitals of the C$_4$ sp-hybridized chain, not considered at the TB level.
  (d) DFT and TB molecular orbitals for the system with unconstrained geometry. 
  }
  \label{fig:TB_DFT_validation}
\end{figure}

\begin{figure}[h!]
    \centering
    \includegraphics[width=\linewidth]{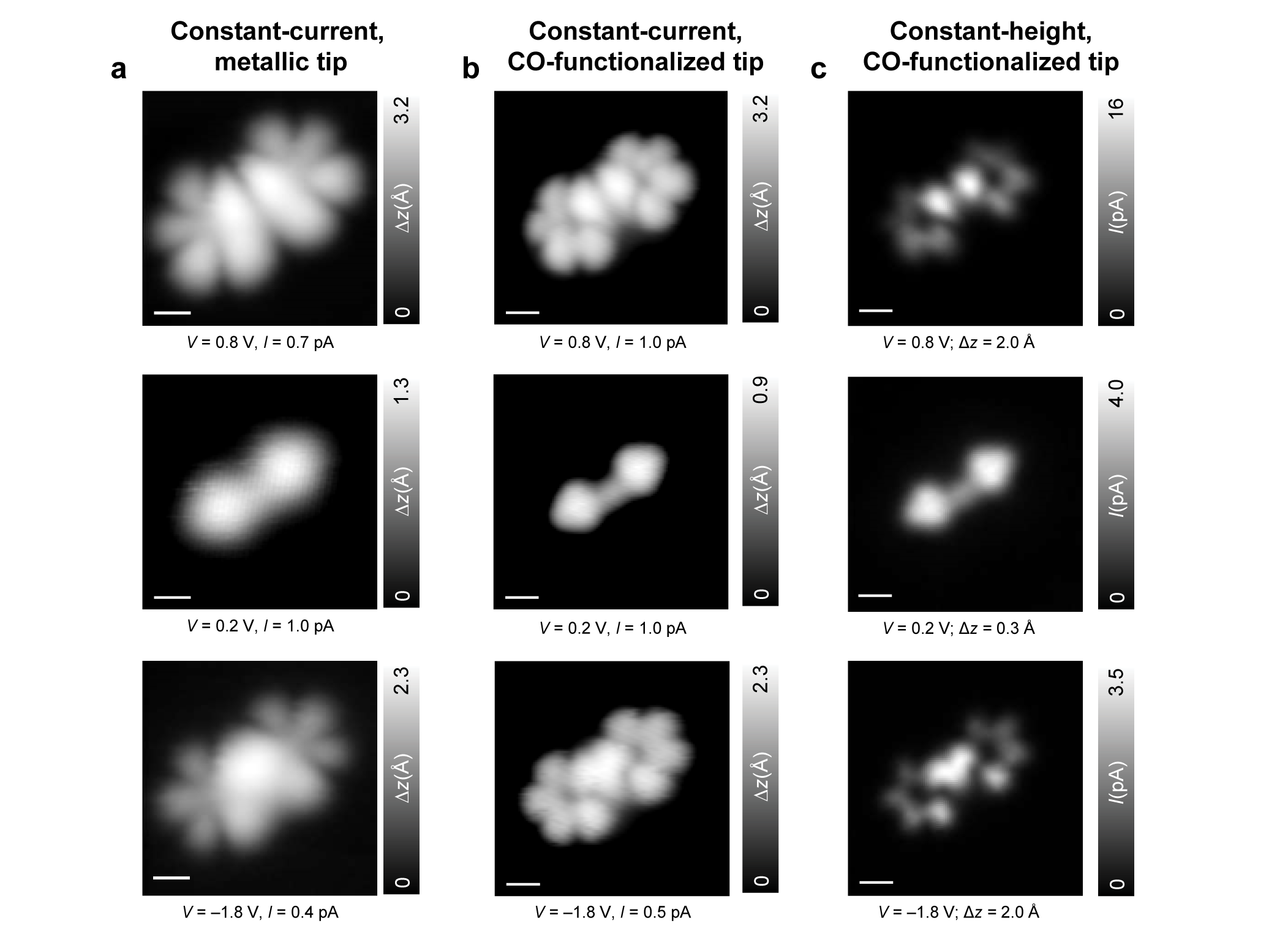}
    \caption{STM images at the negative ion resonance (top), in gap (middle) and positive ion resonance (bottom) of \textit{cis} \textbf{1}, acquired with metallic (a) and CO-functionalized (b, c) tips. The set-point current for constant-height STM images was \textit{I} = 0.5 pA on NaCl, while the set-point voltages were the same as indicated for the respective panels. Scale bars: 0.5 nm.}
    \label{fig:CC CH maps}
\end{figure}

\begin{figure}[h!]
    \centering
    \includegraphics[width=\linewidth]{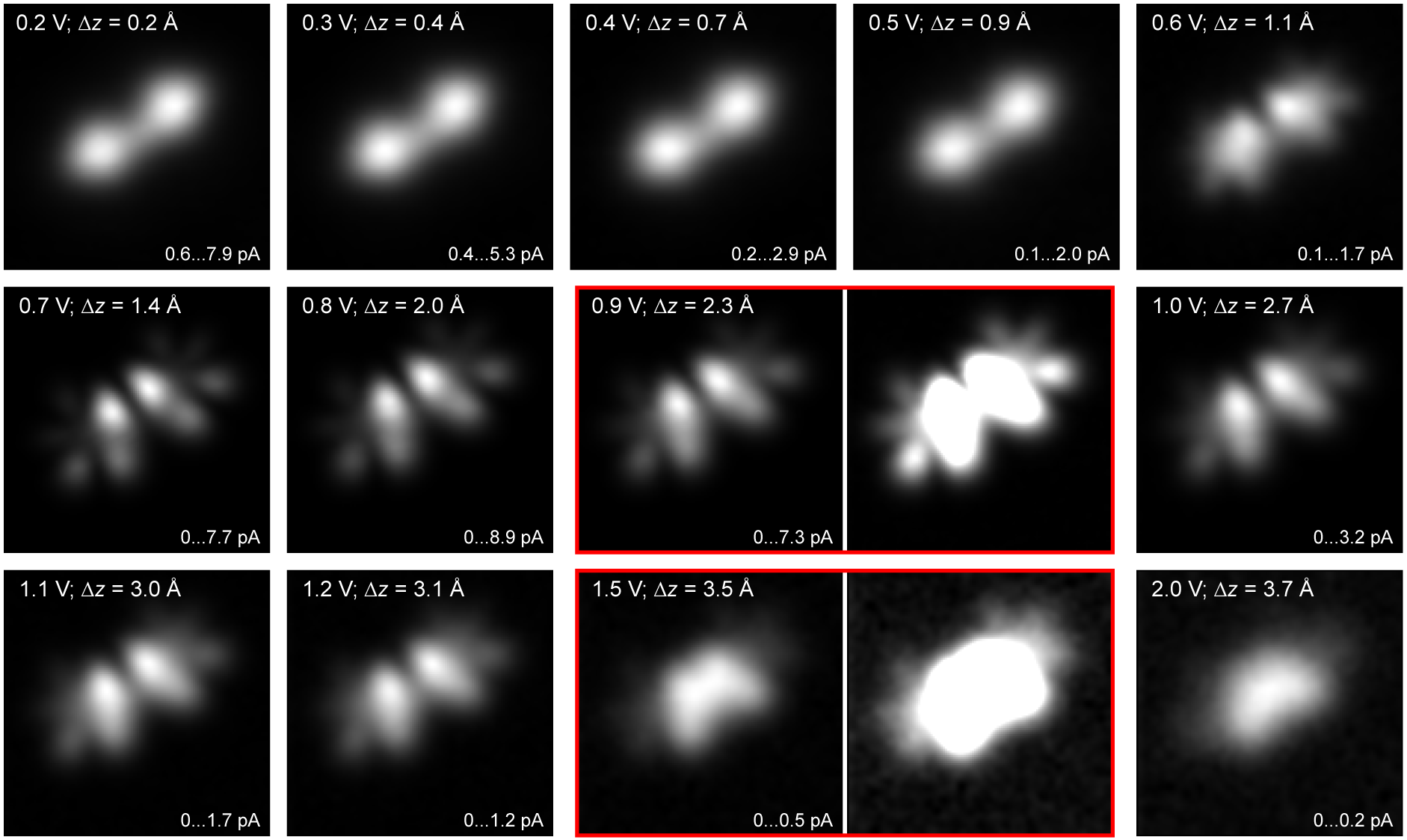}
    \caption{Voltage-dependent STM images of \textit{cis} \textbf{1} at positive bias. STM images at the first two negative ion resonances (\textit{V} = 0.9 and 1.5 V) are highlighted in red. To resolve LDOS features at both the C\textsubscript{4} chain and the phenalenyl units, images at \textit{V} = 0.9 and 1.5 V are shown with two contrast levels. All images were acquired with a metallic tip. The current setpoint was \textit{I} = 0.5 pA on NaCl, while the set-point voltages were the same as indicated for the respective panels. The minimum and maximum values of the current are indicated for each panel. Image sizes: 3.2 nm  $\times$ 3.2 nm.}
    \label{fig:CH maps positive}
\end{figure}

\begin{figure}[h!]
    \centering
    \includegraphics[width=\linewidth]{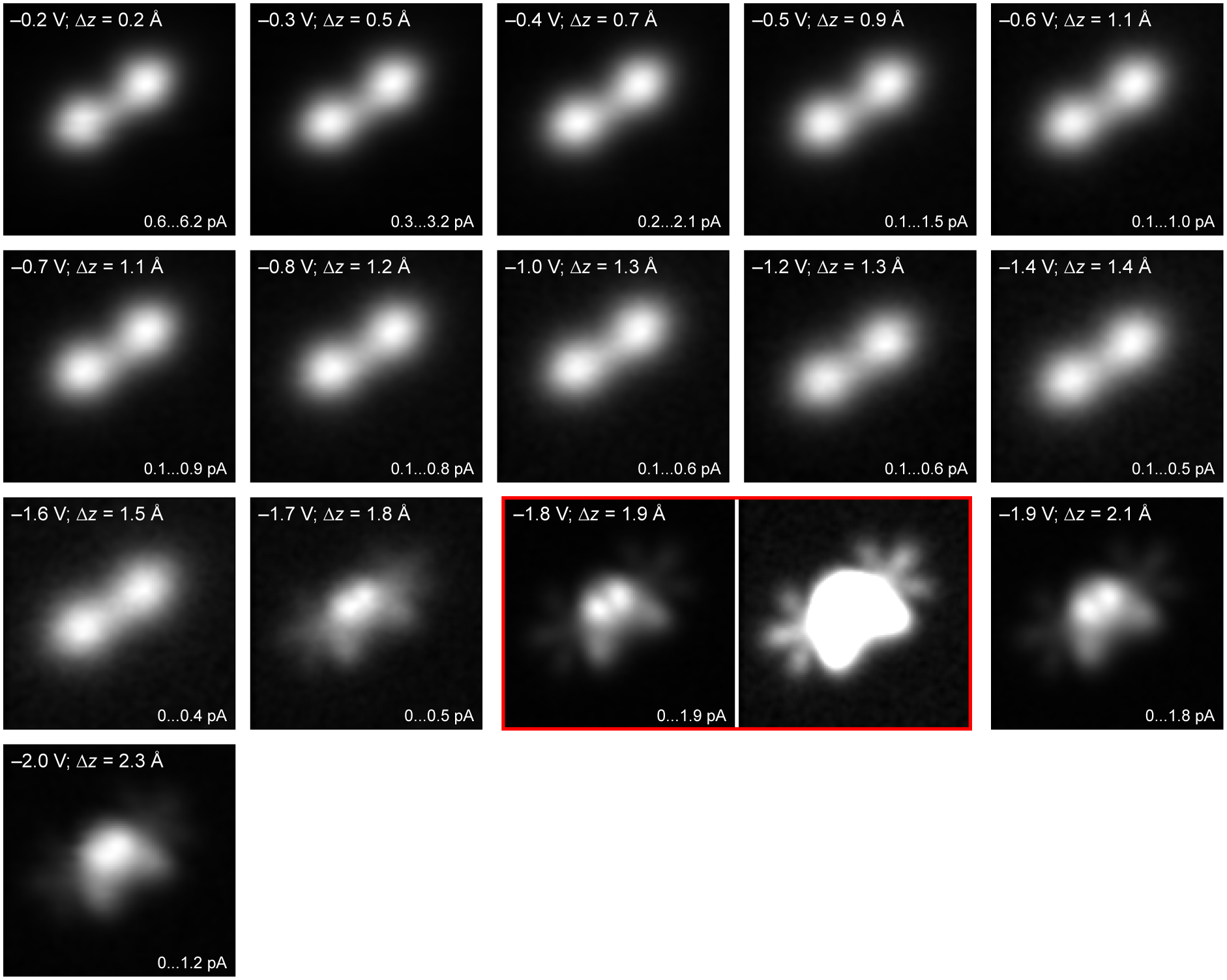}
    \caption{Voltage-dependent STM images of \textit{cis} \textbf{1} at negative bias. STM image at the positive ion resonance (\textit{V} = --1.8 V) is highlighted in red. To resolve LDOS features at both the C\textsubscript{4} chain and the phenalenyl units, the image at \textit{V} = --1.8 V is shown with two contrast levels. All images were acquired with a metallic tip. The set-point current was \textit{I} = 1.0 pA on NaCl, while the set-point voltages were the same as indicated for the respective panels. The minimum and maximum values of the current are indicated for each panel. Image sizes: 3.2 nm  $\times$ 3.2 nm}
    \label{fig:CH maps negative}
\end{figure}
\FloatBarrier

\begin{figure}[h]
    \centering
    \includegraphics[width=\linewidth]{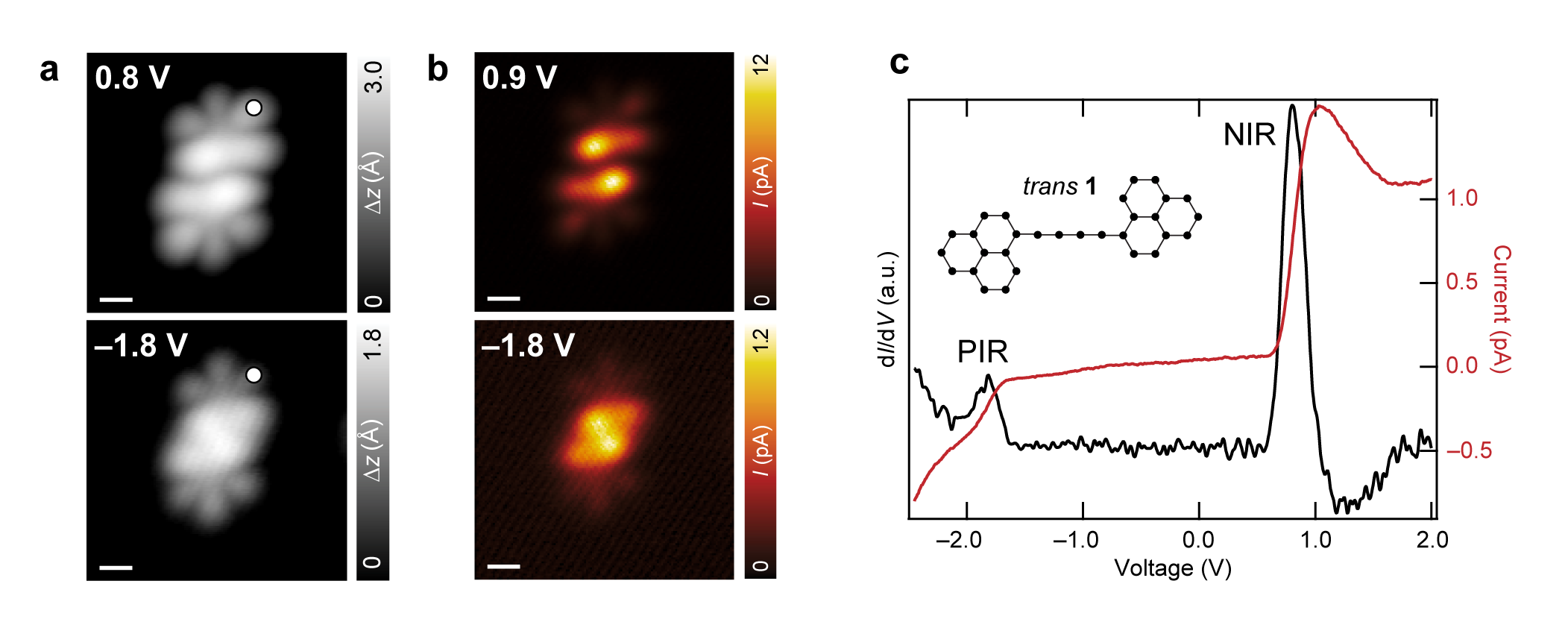}
    \caption{Electronic characterization of \textit{trans} \textbf{1}. (a, b) Constant-current (a) and constant-height (b) STM images of \textit{trans} \textbf{1} at the negative (top) and positive (bottom) ion resonances. Although the STM images of \textit{trans} \textbf{1} appear different from that of the \textit{cis} isomer because of the different molecular geometry, they show the same essential features as observed for the \textit{cis} isomer, namely, a dominant nodal plane at the center of the C\textsubscript{4} chain (0.8 V), and maximum at the center of the C\textsubscript{4} chain along with a weak depression reminiscent of a nodal plane (--1.8 V). The set-point current was \textit{I} = 1.0 pA, with the feedback opened on NaCl for the images in (b). The set-point voltages were the same as indicated for the respective panels. For the images in (b), $\Delta z$ = 2.0 \AA\ (0.9 V) and 1.8 \AA\ (--1.8 V). (c) $I(V)$ spectrum and the corresponding $dI/dV(V)$ spectrum acquired on \textbf{1} at the position indicated by the filled white circles in (a) (open feedback parameters: \textit{V} = --2.5 V, \textit{I} = 1.0 pA). The energetic positions of the positive and negative ion resonances of \textit{trans} \textbf{1} are identical to the \textit{cis} isomer. All images and spectroscopy data were acquired with a metallic tip. Scale bars: 0.5 nm.}
    \label{fig:trans sts}
\end{figure}

\begin{figure}
    \centering
    \includegraphics[width=\linewidth]{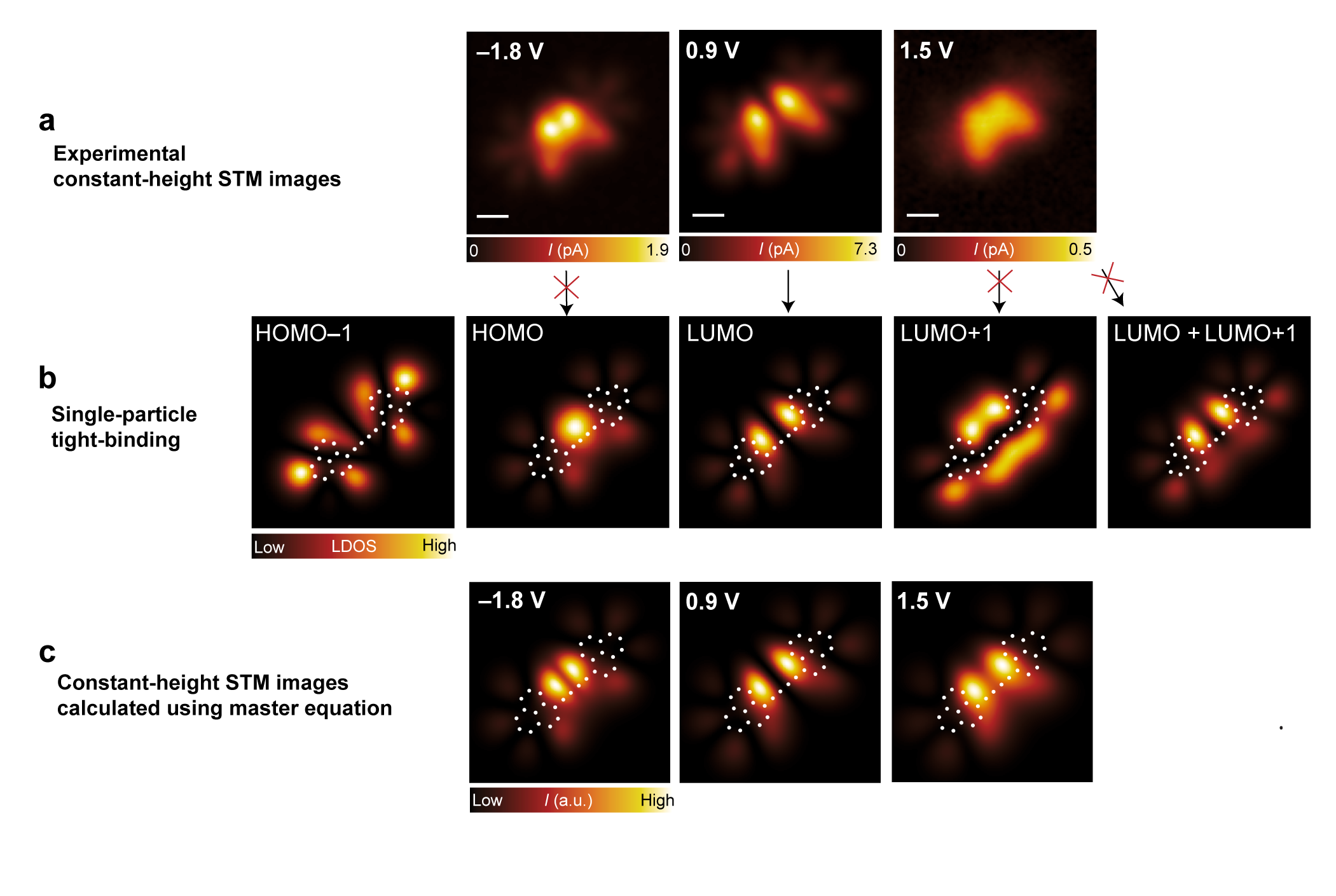}
    \caption{Assignment of STM images at the ion resonances to orbital densities. (a) Experimental constant-height STM images at the ion resonances (reported in the main text). (b) Calculated LDOS maps of the frontier molecular orbitals using TB level of theory ($U=0$ eV). (c) Calculated transition probability maps using master equation at the experimental voltages. Scale bars: 0.5 nm (applies also to the images in panels b and c).}
    \label{fig: failure of single particle}
\end{figure}

\begin{figure}[!htbp]
  \centering
  \includegraphics[width=0.45\columnwidth]{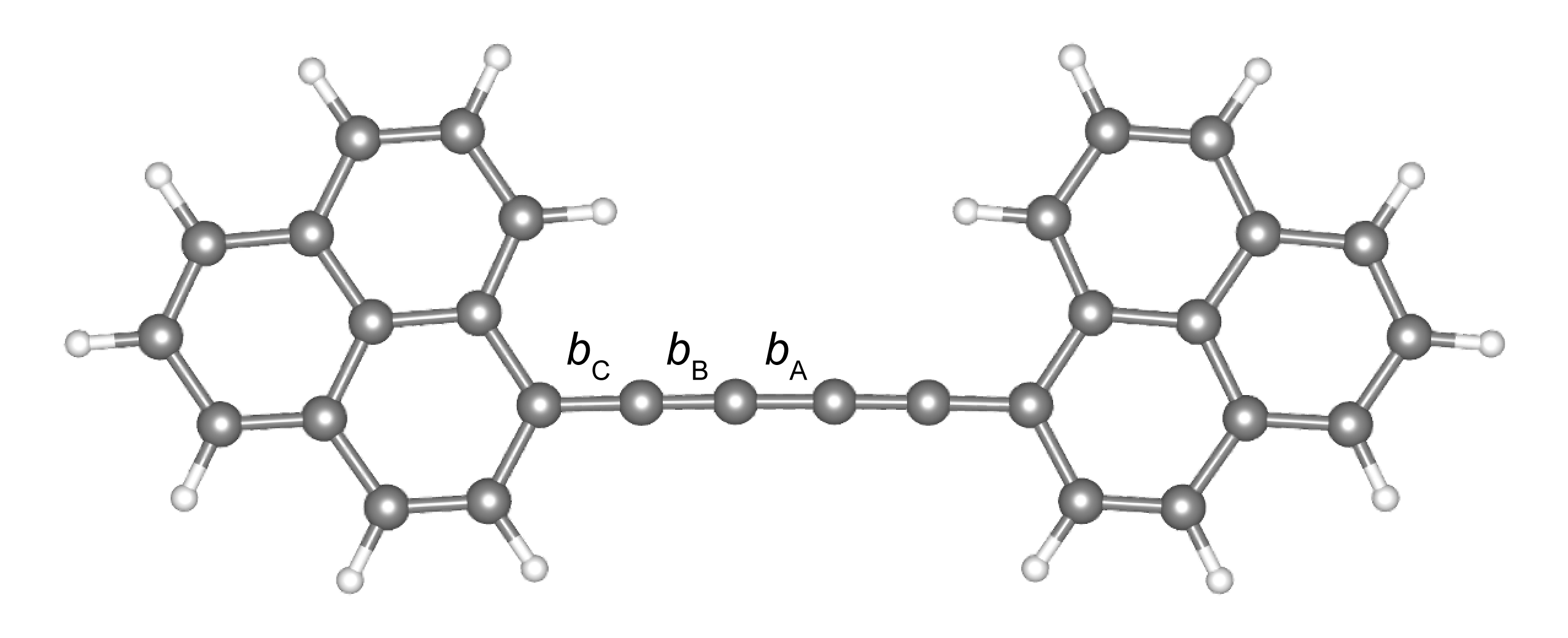}
  \caption{Molecular structure of \textbf{1}. The three bonds $b$\textsubscript{A}--$b$\textsubscript{C} in the central part of the molecule are labeled in accordance with Fig. 1 in the main text.}
  \label{fig:dft_cumulene}
\end{figure}

\begin{figure}
  \centering
  \includegraphics[width=0.6\columnwidth]{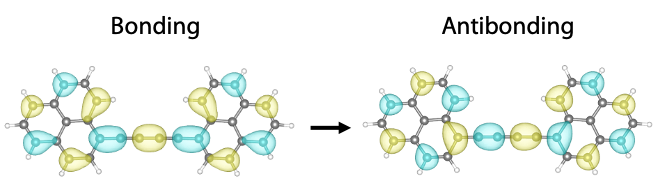}
  \caption{The natural orbitals obtained from SA-CASSCF calculations provide a detailed description of the electronic excitations in both the neutral and charged states. In these states, the primary excitations involve the orbitals with bonding and antibonding symmetries, consistent with the schematic presented in the main text (Fig. 3b).
  }
  \label{fig:no_casscf}
\end{figure}

\begin{figure}[t]
  \centering
  \includegraphics[width=0.5\columnwidth]{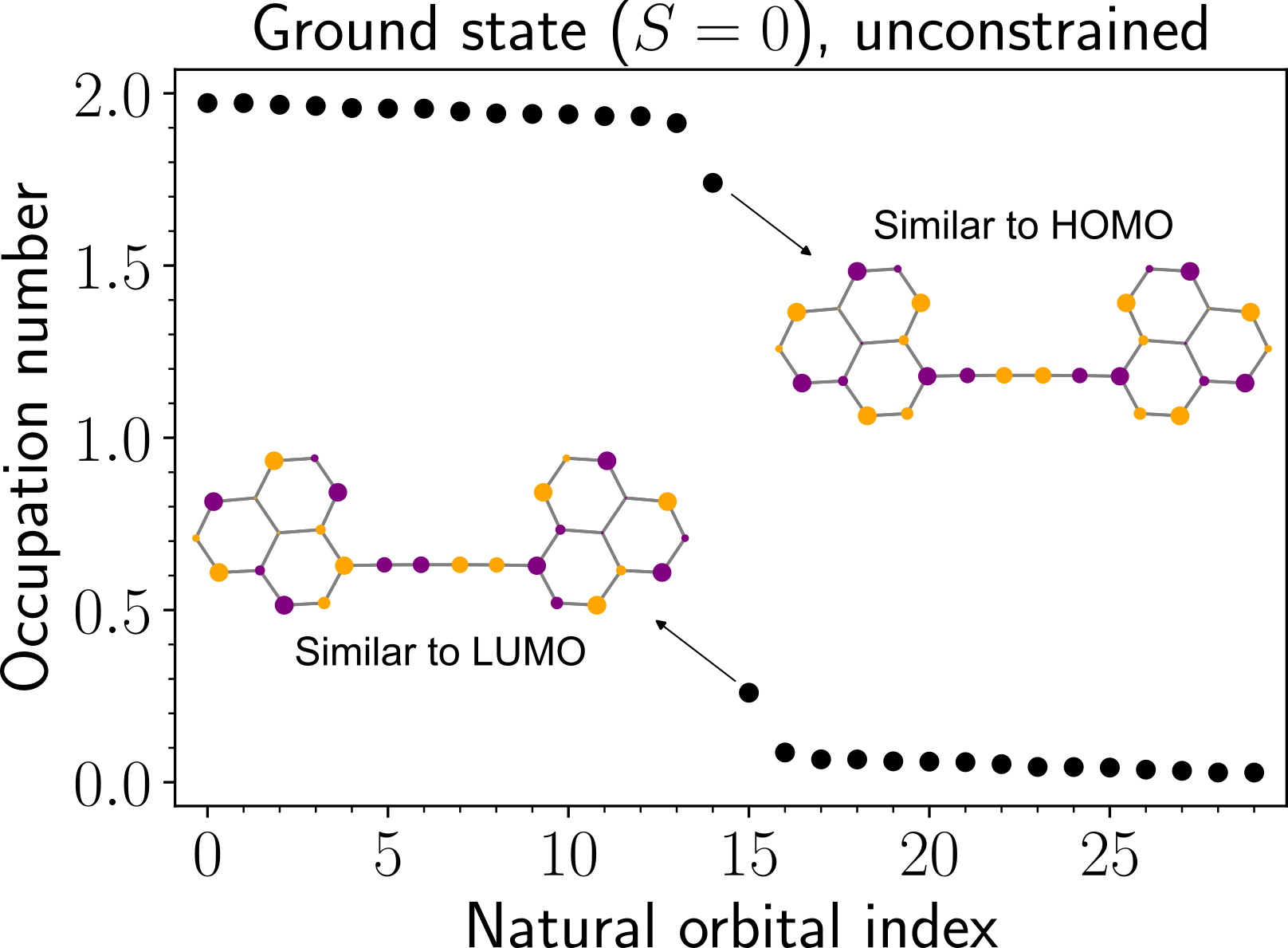}
  \caption{Natural orbital analysis for the ground state of the system with unconstrained geometry.
  Occupation numbers of natural orbitals are computed by DMRG.
  Insets show the natural orbitals with the most fractional occupancies (i.e., those with occupancies furthest from 0 and 2), whose shapes resemble the HOMO and LUMO (cf. Fig.~\ref{fig:TB_DFT_validation}d).}
  \label{fig:NO_analysis}
\end{figure}

\begin{figure}[t]
    \centering
    \includegraphics[width=\linewidth]{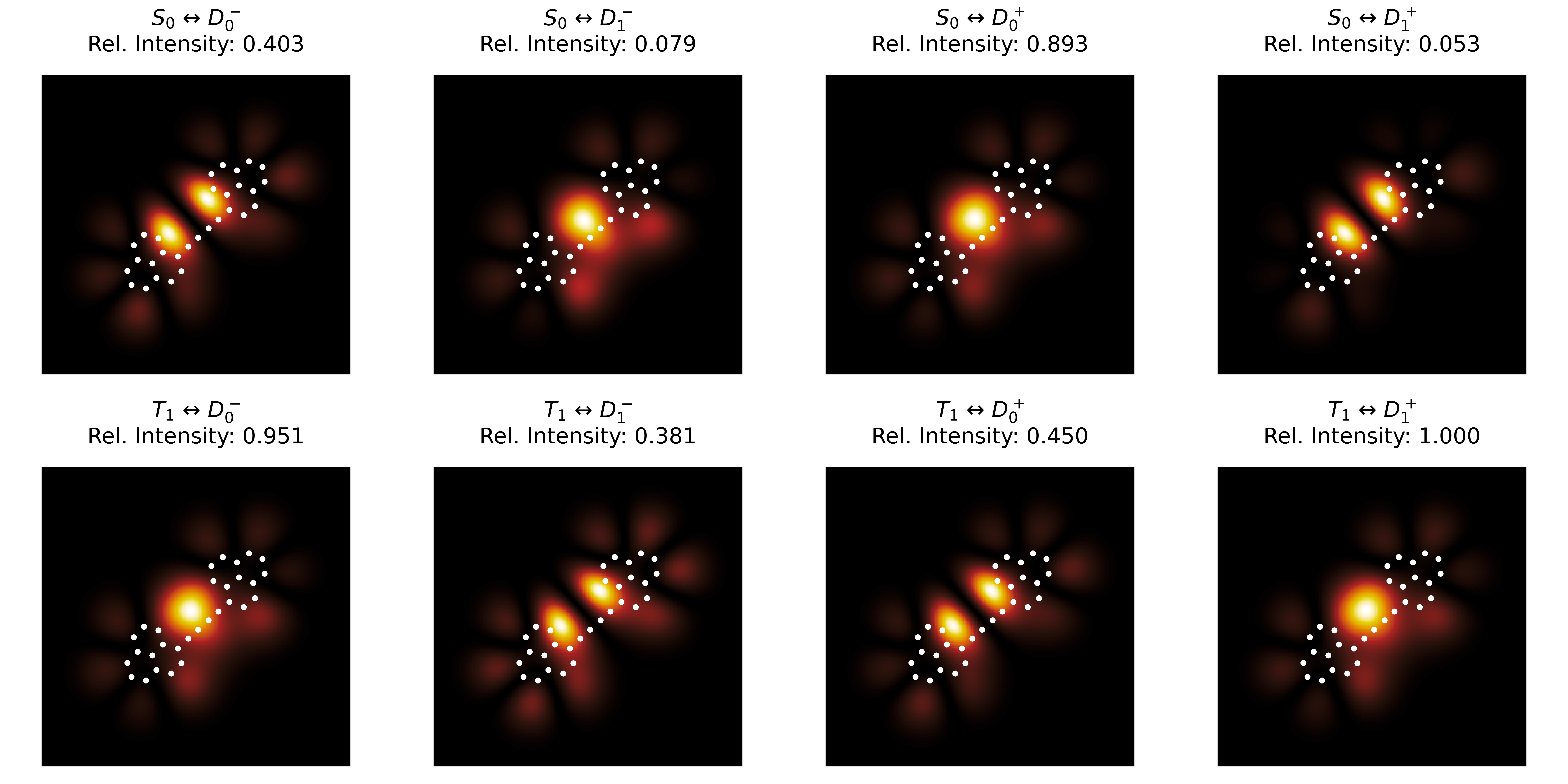}
    \caption{
    Constant-height LDOS maps of Dyson orbitals, corresponding to all transitions between the considered neutral and charged states. The maps were calculated by DMRG for the unconstrained geometry using Slater-type $p_z$ orbitals at a height of $z=20$\,\AA\ above the molecule. The relative intensity is determined by the maximal intensity of each map, divided by the maximum of the $T_1 \leftrightarrow D_1^+$ transition. Image sizes: 2.9 nm  $\times$ 2.8 nm.
    }
    \label{fig:Dyson transitions}
\end{figure}

\begin{figure}[t]
  \centering
  \includegraphics[width=0.8\columnwidth]{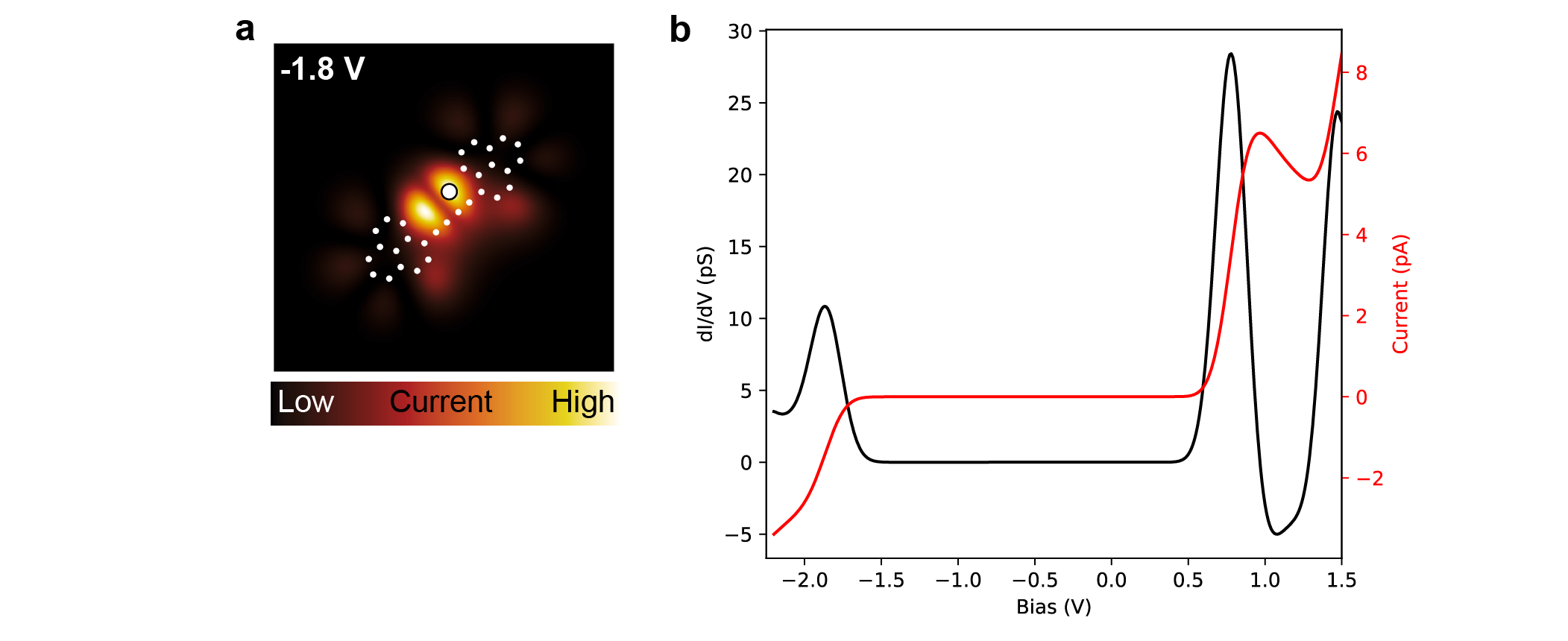}
  \caption{Calculated electronic structure of \textit{cis} \textbf{1} using the master equation.
  (a) Transition probability map at $-1.8$\,V and $z=10$\,\AA.
  (b) \textit{I}(\textit{V}) spectrum and the corresponding d\textit{I}/d\textit{V}(\textit{V}) spectrum for the position indicated by the white filled circle in (a) and at a height of $z=11$\,\AA. Image size: 2.9 nm  $\times$ 2.8 nm.
  }
  \label{fig:ME_spectroscopy}
\end{figure}

\textbf{Failure of the single-particle description.} The constant-height STM images at the PIR and NIR of \textbf{1}, shown in Fig. 3 and Fig. \ref{fig: failure of single particle}a, exhibit a dominant bonding–antibonding character that could be interpreted as the HOMO and LUMO densities of the closed-shell molecule. Using a simple TB approach, the calculated HOMO and LUMO orbitals (Fig. \ref{fig: failure of single particle}b) somewhat agree with the experimentally observed features, apart from the weak depression reminiscent of a nodal plane in the STM image at the PIR, which is attributed to transition involving the neutral triplet state. In addition, a notable discrepancy arises for the NIR+1 measured at 1.5 V: the experimental STM image at the NIR+1 in Fig. \ref{fig: failure of single particle}a does not match either the calculated LUMO+1 LDOS map, or the LDOS map corresponding to a superposition of the LUMO and LUMO+1 in Fig. \ref{fig: failure of single particle}b.

\clearpage

\clearpage
\bibliography{refs_all}